\documentclass[a4paper, amsfonts, amssymb, amsmath, reprint, twocolumn, showkeys, nofootinbib]{revtex4-2}

\usepackage[left=15mm,right=15mm,top=2.5cm,bottom=2cm, columnsep=15pt]{geometry} 

\usepackage{adjustbox}
\usepackage{placeins}
\usepackage[T1]{fontenc}
\usepackage{lipsum}
\usepackage{csquotes}

\usepackage[english]{babel}
\usepackage[utf8]{inputenc}
\usepackage[colorinlistoftodos, color=green!40, prependcaption]{todonotes}
\usepackage[pdftex, pdftitle={Article}, pdfauthor={Author}]{hyperref} 

\usepackage{amssymb}
\usepackage{amsthm}
\input{epsf}
\usepackage{graphicx}
\usepackage{ae,aecompl}

\usepackage{hyperref}
\usepackage{amsmath}
\usepackage{amssymb}
\usepackage{mathtools}
\usepackage{bm}
\usepackage{cleveref}
\usepackage{tensor}
\usepackage{braket}
\usepackage[version=4]{mhchem}
\usepackage{physics}
\usepackage{placeins}

\usepackage{xcolor}
\usepackage{academicons}
\definecolor{orcidlogocol}{HTML}{A6CE39}
\newcommand{\orcid}[1]{\href{https://orcid.org/#1}{\textcolor[HTML]{A6CE39}{\aiOrcid}}}

\usepackage{color}

\def\be{\begin{equation}}
\def\ee{\end{equation}}
\def\beq{\begin{eqnarray}}
\def\eeq{\end{eqnarray}}

\def\apj{{Astrophys. J.}}                 

\def\mnras{{Mon. Not. R. Astron. Soc.}} 
\def\prd{{Phys.~Rev.~D}}        
\makeatletter
\renewcommand{\p@subsection}{}
\renewcommand{\p@subsubsection}{}
\makeatother

\begin{document}

\author{
 L.~Gavassino, M.~Antonelli
}


\vspace{2mm}

    \affiliation{ \vspace{2mm}
        Nicolaus Copernicus Astronomical Center, Polish Academy of Sciences, ul. Bartycka 18, Warsaw, Poland
        \vspace{2mm}
    \\
\vspace{2mm}
\small{ 
 \href{https://orcid.org/0000-0002-6603-9253}{\emph{lorenzo@camk.edu.pl} (L.G.)}, 
 \href{https://orcid.org/0000-0002-5470-4308}{\emph{mantonelli@camk.edu.pl} (M.A.)}
 }
\vspace{2mm}
    }

\title{
Unified Extended Irreversible Thermodynamics and the stability\\ 
of relativistic theories for dissipation
} 
 
 
\begin{abstract}
In a relativistic context, the main purpose of Extended Irreversible Thermodynamics (EIT) is to generalize the principles of non-equilibrium thermodynamics to the domain of fluid dynamics. In particular, the theory aims at modelling any diffusion-type process (like heat as diffusion of energy, viscosity as diffusion of momentum, charge-conductivity as diffusion of particles) directly from thermodynamic laws. Although in Newtonian physics this task can be achieved with a first-order approach to dissipation (i.e. Navier-Stokes-Fourier like equations), in a relativistic framework the relativity of simultaneity poses serious challenges to the first-order methodology, originating instabilities which are, instead, naturally eliminated within EIT. 
The first part of this work is dedicated to reviewing the most recent progress made in understanding the mathematical origin of this instability problem. 
In the second part, we present the formalism that arises by promoting non-equilibrium thermodynamics to a classical effective field theory. We call this approach Unified Extended Irreversible Thermodynamics (UEIT), because it contains, as particular cases, EIT itself, in particular the Israel-Stewart theory and the divergence-type theories, plus Carter's approach and most branches of non-equilibrium thermodynamics, such as relativistic chemistry and radiation hydrodynamics. 
We use this formalism to explain why all these theories are stable by construction (provided that the microscopic input is correct), showing that their (Lyapunov) stability is a direct consequence of the second law of thermodynamics.
\end{abstract}

\maketitle

\section{Introduction} 

In Newtonian physics, the Navier-Stokes equations emerge as an ``almost'' (see \cite{jain_kovtun_2020arXiv}) universal behaviour of dissipative simple fluids when the following double limit is taken \citep{huang_book}: 
\begin{enumerate}
\item[i-] Small deviations from equilibrium.
\item[ii-] Slow macroscopic evolution in an assigned reference frame $A$.
\end{enumerate}
Only when both these conditions are met, more fundamental descriptions (such as kinetic theory or many-body dynamics) can be consistently replaced by the Navier-Stokes equations at the macroscopic scale \cite{KADANOFF1963}. Unfortunately, this same approach in relativity is known to be problematic. For example, the simplest relativistic generalization of the Navier-Stokes equations (the Eckart theory of dissipative fluids) has strong drawbacks, among them the lack of stable solutions and the acausal propagation of perturbations \citep{Hishcock1983,Hiscock_Insatibility_first_order}.  

The requirement (i) that the fluid system is close to equilibrium is a relativistic invariant statement. At the mathematical level, it amounts to a linearization of the equations around a reference state.
This procedure is usually harmless, meaning that the limit in (i) can be safely taken also in a relativistic framework. 
However, the condition (ii) of slow evolution is fundamentally observer-dependent (a slow process in one reference frame may be fast in another reference frame), and involves the potentially dangerous operation of neglecting some higher-order derivatives in time in the reference frame $A$.
In general, this procedure changes the character of a differential equation (typically from hyperbolic to parabolic), so that it can also change the nature of the initial value problem. 
In relativity, this is particularly problematic because, as a consequence of the relativity of simultaneity \citep{special_in_gen}, different observers impose their initial conditions on different time-slices and may, therefore, deal with initial-value problems of different nature \citep{GavassinoLyapunov_2020}. 
The result is that, if we take the slow limit, our final equations (although approximately correct in the technical sense of \citet{Geroch95}) may be impossible to solve in any other reference frame $B \neq A$, as discussed by \citet{Kost2000}. 
This is due to the fact that in $B$ any initial condition, given with a finite precision, might unavoidably contain fast-growing  modes that are non physical \citep{Hiscock_Insatibility_first_order}.

The solution proposed by \citet{Israel_Stewart_1979} consists of dropping the slow limit assumption (ii) and working only under the assumption of small deviations from equilibrium. In fact, their theory includes a consistent description of the (possibly fast) relaxation processes of the fluid elements towards local thermodynamic equilibrium. The causal and stable thermodynamics of Israel and Stewart provides a satisfactory replacement of the unstable and non-causal models of Eckart and Landau \& Lifshitz: in the Israel-Stewart framework, the observer-dependent slow limit can be avoided altogether and the equations keep a well defined hyperbolic structure. The structure of the Israel and Stewart's theory can be seen as the prototype for a general relativistic formulation of \emph{Extended Irreversible Thermodynamics} (EIT) \citep{Jou_Extended,rezzolla_book}. 
 
Although the Israel-Stewart theory has satisfactory mathematical properties \citep{Hishcock1983} and is in good agreement with kinetic theory \citep{Israel_Stewart_1979}, it fails to be a universal theory \citep{Salazar2020}. The reason is that the slow limit (ii) was a necessary element to extract a universal behaviour out of the linearized equations, which may otherwise exhibit exotic dynamics \citep{Denicol2012Boltzmann,Heller2014} and may posses an infinite number of degrees of freedom \citep{Denicol_Relaxation_2011}. 
As an example, even in the simple case of pure bulk viscosity, there is an infinitely large class of systems (consistently modelled using non-equilibrium thermodynamics) that cannot be described within the Israel-Stewart theory \citep{BulkGavassino}. All this is discussed with the aid of two simple examples (the Klein-Gordon field and photon diffusion) in Sections 2-4, and more in general in Section 5.

The aim of the second part of this work (starting with Section 6) is to review and extend the approach of Israel and Stewart by proposing a formalism that might have a more universal applicability (if not as a precise limit, at least as a useful approximation) and which constitutes a natural hydrodynamic extension of non-equilibrium thermodynamics. We will refer to this approach as \textit{Unified Extended Irreversible Thermodynamics} (UEIT), since it unites within a single formalism the relativistic EIT theories (like the Israel-Stewart model and the so-called divergence-type theories, see \citet{Liu1986}) with Carter's variational approach \citep{carter1991,Carter_starting_point}, radiation hydrodynamics \citep{mihalas_book} and relativistic chemistry \citep{noto_rel}. 
The assumptions, and the associated limitations, of UEIT will be discussed as well.

The main goal of UEIT is not just to provide a well-behaved formulation of relativistic dissipative hydrodynamics \cite{AnilePavon1998}, a task that seems to be already achieved by the so-called \emph{frame-stabilised} first-order theories \citep{Kovtun2019,BemficaDNDefinitivo2020}, but also to embed the principles of non-equilibrium thermodynamics in an curved space-time. 
In fact, contrarily to what is done in first-order theories, UEIT overcomes the instability problem of the relativistic Navier-Stokes approach by means of the same, fundamental, thermodynamic principles that also guarantee the (Lyapunov) stability of any other dissipative system \citep{GavassinoLyapunov_2020,Prigogine1978,PrigoginebookModernThermodynamics2014}. 
Therefore,  UEIT should not be considered as a specific hydrodynamic theory, but rather as a thermodynamic language, that can be used to formulate a class of classical effective field theories which are connected with statistical mechanics in a natural way.  

The ideas at the basis of UEIT have been around for a long time, without being systematized in a global picture. For this reason, our presentation will follow the structure of an introductory review, collecting together some important results on the topic. Formal aspects will be addressed with the aid of simplified physical models, possibly referring to the original research works for the formal and more technical details. 

Our approach to the subject is complementary to that of most of the other reviews on EIT, like \citet{Jou_Extended,Salazar2020}. In fact, the discussion usually revolves around the thermodynamic foundations of the theory and on the principles of transient thermodynamics that lead to EIT (e.g., the \textit{release of variation} postulate). 
However, since hydrodynamics is a phenomenological description that finds application in diverse areas of physics (ranging from heavy-ion collisions to neutron star physics),  EIT has evolved in each area autonomously,  partially losing its original connection with the thermodynamic principles which led to its formulation.
For example, in dilute-gas physics EIT is regarded as an approximation to the Boltzmann equation \citep{Denicol2012Boltzmann,chabanov2021}, whereas in dense-matter physics EIT is a synonym of multi-fluid modelling \citep{BulkGavassino,RauWasserman2020}. For this reason, in this review we will privilege an effective field theory perspective, which, we believe, clarifies the rationale of EIT and makes its physical content more accessible. 
\\
\\
As our main interest is to discuss the thermodynamic interpretation of various models, the spacetime (not necessarily flat in some sections) will be treated as a fixed background, so we will not include the metric in the set of the dynamical fields of the theory and the possible dependence of some quantity on $g_{\nu \rho}$ will be tacit. We adopt the spacetime signature $ ( - , +, + , + ) $ and work in natural units $c=\hbar =k_B=1$. The space-time indices $\nu,\rho,\sigma...$ run from 0 to 3, while the indices $j,k$ are always space indices and run only from 1 to 3. The indices $i,h$ are always abstract labels, counting the fields of a specific hydrodynamic theory.

\section{Hydrodynamics as a classical field theory}

To set terminology and notation, in this section we provide a minimal introduction to relativistic hydrodynamics  interpreted as a classical effective field theory  \citep{Dubosky2012_effective,kovtun_lectures_2012}.

\subsection{Constitutive relations and hydrodynamic equations}\label{iPPhi}

The bold bet of the hydrodynamic point of view is that it is possible to use a limited set of macroscopic independent (but possibly subject to some algebraic constraints) classical fields $\varphi_i$  on the spacetime manifold $\mathcal{M}$ to describe the large-scale evolution of multi-particle systems ($i$ is an index labelling the fields). 
Such a description is useful if it enables to calculate the values of 
\begin{enumerate}
\item the (symmetric) stress-energy tensor $T^{\nu \rho}$ of the fluid system,
\item the entropy current $s^\nu$,
\item any conserved current
\end{enumerate}  
at every point of $\mathcal{M}$, see  \citep{Israel_2009_inbook}. 
Therefore, once the choice of the fields $\varphi_i$ is made, one needs to provide some formulas 
\begin{equation}
\label{functional}
\begin{split}
& T^{\nu \rho} = T^{\nu \rho}(\varphi_i,\nabla_\sigma \varphi_i, \nabla_\sigma \nabla_\lambda \varphi_i,...) \\
& s^\nu = s^\nu (\varphi_i,\nabla_\sigma \varphi_i, \nabla_\sigma \nabla_\lambda \varphi_i,...) \\
& n^\nu = n^\nu (\varphi_i,\nabla_\sigma \varphi_i, \nabla_\sigma \nabla_\lambda \varphi_i,...) \, , 
\end{split}
\end{equation}
which are called \textit{constitutive relations} \citep{Liu1986,Kovtun2019,kovtun_lectures_2012}. 
For simplicity, we are assuming that there is only one independent conserved particle (minus anti-particle) current $n^\nu$, meaning that the fluid is a \textit{simple} fluid \citep{MTW_book}. In writing \eqref{functional} we have implicitly made use of a locality assumption, namely the physical tensors at an event depend only on the fields and their covariant derivatives evaluated at the same event. 

To complete the hydrodynamic model, we need to specify the evolution of the fields $\varphi_i$ by prescribing a set of differential \textit{hydrodynamic equations}
\begin{equation}
\label{hydrouzZZ}
\mathfrak{F}_h(\varphi_i,\nabla_\sigma \varphi_i, \nabla_\sigma \nabla_\lambda \varphi_i,...)=0 \, . 
\end{equation} 
The quantities $\mathfrak{F}_h$ are tensor fields and, again, the same locality assumption has been invoked.
 
The constitutive relations \eqref{functional} and the hydrodynamic equations \eqref{hydrouzZZ} must be formulated in such a way to ensure the validity of the fundamental conservation laws
\begin{equation}
\label{conservotutto}
\nabla_\nu T^{\nu \rho}=0 \quad   \quad \quad    \nabla_\nu n^\nu =0 \, ,
\end{equation} 
and of the second law of thermodynamics
\begin{equation}
\label{secondlaw}
\nabla_\nu s^\nu \geq 0 \, ,
\end{equation}
at least to the level of accuracy that the theory is designed to have. A hydrodynamic theory is said to be \emph{non-dissipative} when the entropy production \eqref{secondlaw} is exactly zero, while it is said to be \emph{dissipative} otherwise. 

Finally,  any hydrodynamic model should (in principle) respect some compatibility constraints with
\begin{enumerate}
\item[i-] \textit{Equilibrium statistical mechanics}: isolated dissipative systems should eventually converge (for $t \longrightarrow +\infty$) to a global thermodynamic equilibrium state, whose properties should coincide with those computed by means of the micro-canonical ensemble \citep{huang_book}. Therefore, this state must be stable against any perturbation allowed by the hydrodynamic model. In non-dissipative theories this thermodynamic equilibrium state still exists and is stable under perturbations (i.e. the system shall not spontaneously depart from it), but, since entropy is conserved, the system does not converge to it for large times. More formally, we may say that both dissipative and non-dissipative systems are Lyapunov stable, but only dissipative ones are also asymptotically stable.
\item[ii-] \textit{Kinetic theory}: if the system admits a kinetic description, then the predictions of the hydrodynamic theory (within its range of applicability) should coincide with those of kinetic theory. 
As a consequence, the possible presence of exact kinematic constraints, such as the traceless condition for the stress-energy tensor of ultra-relativistic ideal gases ($T\indices{^\nu _\nu}=0$), need to be carefully encoded into  the hydrodynamic model. 
\item[iii-] \textit{Causality}: the fundamental principles of Quantum Field Theory (namely, equal-time commutation/anticommutation relations and Lorentz covariance) impose that information cannot propagate faster than light \citep{Peskin_book}. Therefore, this is a property of every physical system that should be respected by phenomenological models of the kind we are describing here 
(within the precision and physical limits of the model). 
\end{enumerate}
While a hydrodynamic model may, in principle, implement the requirements (ii) and (iii) only in an approximate way, we will see that (i) is of central importance for UEIT, especially if the model is  dissipative. In fact, the purpose of a dissipative model is to describe how a system spontaneously evolves towards thermodynamic equilibrium by converting the energy of the macroscopic hydrodynamic motion into internal energy.

\subsection{Setting the terminology: perfect fluids}\label{perflabgeulftssa}

To gain some initial insights, let us discuss a specific implementation of \eqref{functional} and \eqref{hydrouzZZ}. 
To set the terminology, let us consider the perfect fluid. The most common way to formulate its dynamical equations is within the Eulerian specification of the flow over $\mathcal{M}$; within this Eulerian framework, the fields can be chosen as
\begin{equation}
\label{perf_flu_fields}
(\varphi_i) \, =  \, ( \, u^\sigma   , \, s \, , \, n \, ) \, ,
\end{equation} 
where  the vector field $u^\sigma$ is interpreted as the fluid four-velocity, while the scalar fields $s$ and $n$ are respectively the local entropy and particle density, as measured in the reference frame defined by $u^\sigma$,
\begin{equation}\label{tuttiCampi}
 u^\sigma =  {n^\sigma}/{\sqrt{-n_\lambda n^\lambda}} 
\quad \quad 
s=-s^\sigma u_\sigma  
\quad \quad
n=-n^\sigma u_\sigma \, .
\end{equation}
Given the choice in \eqref{perf_flu_fields}, the constitutive relations \eqref{functional} read
\begin{align}
    \label{functional333}
    \begin{split}
& T^{\nu \rho} = (\mathcal{U}+\Psi)u^\nu u^\rho + \Psi g^{\nu \rho} 
 \\ 
& s^\nu = s u^\nu 
 \\
& n^\nu = n u^\nu \, . 
\end{split}
\end{align}
The quantities $\mathcal{U}$ and $\Psi$ are respectively the energy density and the pressure of the fluid, as measured in the frame of $u^\sigma$. The constitutive relation for the stress-energy tensor is completed by an equation of state $\mathcal{U}=\mathcal{U}(s,n)$, whose differential is
\begin{equation}
\label{EEEEOOS}
d\mathcal{U} = \Theta ds +\mu dn \, .
\end{equation}
The pressure $\Psi$ is related to the energy density through the Euler relation\footnote{Similarly to what is done in non-relativistic thermodynamics, the Euler relation is often derived from an additivity property of the system, see \citep{andersson2007review} or \citep{Termo} for the more general case of perfect multifluids.}, 
\begin{equation}\label{EULRELO}
\mathcal{U} + \Psi = \Theta s + \mu n \, .
\end{equation} 
Finally, the hydrodynamic equations \eqref{hydrouzZZ} read
\begin{equation}
\label{hydroequez}
\begin{split}
& (\mathcal{U}+\Psi) u^\nu \nabla_\nu u_\rho +(\delta\indices{^\nu _\rho}+u^\nu u_\rho) \nabla_\nu \Psi =0 \\
& u^\nu \nabla_\nu s + s \nabla_\nu u^\nu =0 \\
& u^\nu \nabla_\nu n + n \nabla_\nu u^\nu =0 \, .  
\end{split}
\end{equation}
Combining the constitutive relations together with the hydrodynamic equations immediately leads to \eqref{conservotutto} and \eqref{secondlaw}, with~$\nabla_\nu s^\nu =0$. 

The same system can be described by using two arbitrary  independent thermodynamic variables as fundamental fields \citep{MTW_book}. 
For example, an alternative to \eqref{perf_flu_fields} could be \citep{Kovtun2019}
\begin{equation}\label{kovtunN}
(\varphi_i) \, = \, ( \,  u^\sigma, \,  \Theta\,  ,\,  \mu\,  ) \,  .
\end{equation}
The two descriptions are related by a change of variables, where $s$ and $n$ should be written in terms of $\Theta$ and $\mu$. This can be conveniently done by trying to obtain an expression of the kind $\Psi = \Psi (\Theta,\mu)$. 
In fact, from \eqref{EEEEOOS} and \eqref{EULRELO}, one can derive the differential
\begin{equation}
\label{JonasIlBriccone}
d\Psi \, = \, s d\Theta + n d\mu \, ,
\end{equation}
which immediately provides the relations $s(\Theta,\mu)$ and~$n(\Theta,\mu)$.

Perfect-fluid dynamics can also be obtained starting from a different perspective (as opposed to the Eulerian point of view), namely via the Lagrangian specification of the flow. In this approach, also known as pull-back formalism \citep{andersson2007review}, the fields of the theory are three scalar fields,
\begin{equation}\label{LagrangianFields}
(\varphi_i) \,=\, (\,\Phi^1,\Phi^2,\Phi^3\,)\, ,
\end{equation} 
representing the comoving Lagrangian coordinates of the fluid elements \citep{Comer_1993,prix2004,ballesteros_prd2016,Termo}. 
These fields are sometimes referred to as matter-space coordinates \citep{andersson2007review}, and are related to the fluid velocity $u^\nu$ via
\begin{equation}
u^\nu \nabla_\nu \Phi^1 = u^\nu \nabla_\nu \Phi^2 =u^\nu \nabla_\nu \Phi^3 =0 \, .
\end{equation}
These conditions are automatically satisfied if the constitutive relations for the currents have the form \begin{equation}\label{mictution}
\begin{split}
& s^\nu =\varepsilon^{\nu \rho \sigma \lambda} \, S(\Phi^1,\Phi^2,\Phi^3) \, \nabla_\rho \Phi^1 \nabla_\sigma \Phi^2 \nabla_\lambda \Phi^3 \\
& n^\nu = \varepsilon^{\nu \rho \sigma \lambda} \, N(\Phi^1,\Phi^2,\Phi^3) \, \nabla_\rho \Phi^1 \nabla_\sigma \Phi^2 \nabla_\lambda \Phi^3 \, , 
\end{split}
\end{equation}
where $S$ and $N$ are two arbitrary functions, to be chosen to reproduce the correct initial conditions \citep{andersson2007review}. 
Equations \eqref{mictution} provide a  mapping between the Eulerian and the Lagrangian descriptions: when inserted into the right-hand sides of the equations in  \eqref{tuttiCampi}, we obtain the rules for the change of variables 
\begin{equation}\label{ggrungoGringo}
(u^\sigma,s,n) \longrightarrow (\Phi^1,\Phi^2,\Phi^3) \, .
\end{equation}
It is interesting to note that, while in the Eulerian framework the conservation of $s^\nu$ and $n^\nu$ is imposed at the level of the hydrodynamic equations \eqref{hydroequez}, in the Lagrangian one they are already implemented within the constitutive relations. This means that they are satisfied both \textit{on-shell} (i.e. by fluid configurations that are solutions to the hydrodynamic equations) and \textit{off-shell} (i.e. on a generic spacetime configuration of the fields $\Phi^1,\Phi^2,\Phi^3$), see e.g. \citet{Comer_1993,Termo}. 

An alternative formulation of the Lagrangian approach  has been proposed by \citet{Dubosky2012_effective}, which can be shown to be equivalent to the one summarised here \citep{Termo}. Moreover, the Lagrangian point of view finds many applications in multifluid hydrodynamics \citep{Carter_starting_point,andersson2007review}, elasticity theory \citep{CarterQuintana1972,ander_elasticity_2019}, and models for dissipation \citep{Grozdanov2013,AnderssonComerDissAct2015,TorrieriIS2016,CeloraNils2020}.

\subsection{Non-interacting scalar field} 

For later convenience, we also introduce another minimal example, that is intimately different from the perfect fluid: the complex scalar field, 
\begin{equation}
(\varphi_i) = (\varphi) \, .
\label{qwertyqwerty}
\end{equation}
For simplicity, we assume flat spacetime with global inertial coordinates $x^\nu$ and impose that the dynamics is governed by the Klein-Gordon action \citep{weinbergQFT_1995}
\begin{equation}\label{action}
\mathcal{I}[\varphi] = \int_{\mathcal{M}} \big[ -\partial_\nu \bar{\varphi} \, \partial^\nu \varphi - m^2 \bar{\varphi} \varphi  \, \big]  \, d^4 x \, .
\end{equation}    
For what concerns us, this (phenomenological) model defines the \textit{classical}  dynamics of a macroscopic complex field with mass $m>0$, which is often used (for theoretical purposes) as the minimal example of an unconventional kind of fluid, whose equation of state is a differential equation \citep{KGgeon}. 
When coupled with gravity, this theory is also used to model  boson stars\footnote{
    In fact, the free Klein-Gordon classical field $\varphi$ can be interpreted as the order parameter of a zero-temperature condensate of (strictly) non-interacting scalar bosons of mass $m$, namely 
    $\varphi(x) \sim \bra{N} \hat{\varphi}(x) \ket{N+1}$, see \citep{landau9,CosmoCondesnato1993,Chavanis2012}. }
\citep{Liebling2012}.
The constitutive relations \eqref{functional} can be computed directly from the action principle:
\begin{align}\label{defiziokleim}
\begin{split}
& T^{\nu \rho} = 2 \, \partial^{(\nu} \bar{\varphi} \, \partial^{\rho )} \varphi - g^{\mu \nu} \big( \partial_\alpha  \bar{\varphi} \, \partial^\alpha \varphi +  m^2 \bar{\varphi} \varphi \big) 
 \\
&  s^\nu =0 
  \\
&  n^\nu = i (\varphi \, \partial^\nu \bar{\varphi} - \bar{\varphi} \, \partial^\nu \varphi) \, ,   
  \end{split}
 \end{align}
where $ T^{\nu \rho}$ is the Belinfante-Rosenfeld tensor of the model, while $n^\nu$ is the Noether current associated with the $U(1)$ symmetry of the action. We imposed the condition $s^\nu =0$ as a zero-temperature assumption (the model is non-dissipative). 
The hydrodynamic equation \eqref{hydrouzZZ} is just the usual Klein-Gordon equation obtained from the action \eqref{action},
\begin{equation}\label{klinGordon}
\partial_\nu \partial^\nu \varphi = m^2 \varphi \, ,
\end{equation}
implying that the conservation laws \eqref{conservotutto} are ensured on-shell by construction. 
The solution of \eqref{klinGordon} reads
\begin{equation}\label{soluzionediklein}
\varphi(x) = \int \bigg[ a_p e^{i p_\nu x^\nu} + \bar{b}_p e^{-i p_\nu x^\nu} \bigg] 
\dfrac{dp^1 dp^2 dp^3}{(2\pi)^3\sqrt{2p^0}} \, ,
\end{equation}
where the four-momenta $p^\nu$ must satisfy the constraints
\begin{equation}
\label{inreaLZE}
p^\nu p_\nu =-m^2 \, ,  \, \,    p^0 >0 
 \, \,  \Rightarrow \, \, p^0 = \sqrt{m^2 + p^j p_j}  
\end{equation}
Since every small perturbation can be decomposed into modes with real frequencies, any solution is stable under small perturbations. In addition, the dispersion relation \eqref{inreaLZE} imposes a subluminal signal propagation, ensuring causality.

\subsection{Hydrodynamic degrees of freedom: the scalar field and perfect fluid examples} \label{vrif} 

To introduce the fundamental concept of hydrodynamic degree of freedom,
we start by considering the minimal Klein-Gordon model introduced in the previous subsection, as it builds on the single field in \eqref{qwertyqwerty}. All the relevant information about the system is contained in a single function  $\varphi : \mathcal{M} \longrightarrow \mathbb{C} $, 
meaning that every physical quantity can be computed at each point of $\mathcal{M}$ directly from any given  solution of the form \eqref{soluzionediklein}.

A standard hydrodynamic problem is, however, typically formulated as an initial value problem, where one imposes a set of initial conditions on a  space-like 3D Cauchy hypersurface $\Sigma$ (e.g. the surface $t=0$). 
Then, the evolution of the hydrodynamic quantities in the causal future of $\Sigma$ (e.g. the portion of the spacetime $t>0$) is studied. The number of independent real functions which must be assigned on $\Sigma$ to identify a unique solution is the number of \emph{hydrodynamic degrees of freedom} $\mathfrak{D}$ of the model.

Since equation \eqref{klinGordon} is of the second order in time, the knowledge of the initial condition
\begin{equation}\label{intialCond}
\varphi : \Sigma \longrightarrow \mathbb{C}
\end{equation}
is not enough to uniquely evolve the field configuration. However, 
it is possible to recast the evolution as two complex first-order equations of the Hamilton type,
\begin{equation}
 \partial_t \varphi = \bar{\Pi}_{KG} 
 \quad \quad \quad \quad 
 \partial_t \bar{\Pi}_{KG} = \partial_j \partial^j \varphi -m^2 \varphi \, , 
\end{equation}
where the conjugate momentum is
\begin{equation}
\Pi_{KG} := \dfrac{\delta \mathcal{I}}{\delta (\partial_t \varphi)} = \partial_t \bar{\varphi} \, .
\end{equation}
Now, if together with \eqref{intialCond} we prescribe also the initial conditions of the conjugate momentum,
\begin{equation}
\Pi_{KG} : \Sigma \longrightarrow \mathbb{C} \, ,
\end{equation}
we have a well-posed initial value problem. Hence, the Klein-Gordon model has $\mathfrak{D}=4$ degrees of freedom (i.e. two complex functions). We remarked this basic and well know fact to stress the idea that, in general, the number of degrees of freedom  differs from the number of (real) algebraically independent components of the fields $\varphi_i$. The difference between the two typically depends on the order of the hydrodynamic equations, on the possible presence of gauge freedoms, and on the possibility of some equations to play the role of constraints on the initial conditions (e.g. as in the well known case of Maxwell equations).

For the case of perfect fluids in the Eulerian framework, we have 5 first-order independent dynamical equations \eqref{hydroequez} for 5 variables \eqref{perf_flu_fields}, which constitute a closed system with $\mathfrak{D}=5$. It is interesting to note that, while within the Eulerian specification of the flow field the number of independent components of the fields coincides with the number of degrees of freedom, this is no longer the case when we move to the Lagrangian point of view. In fact, $\Phi^1,\Phi^2,\Phi^3$ are 3 real scalar fields, obeying 3 second-order differential equations, which can be obtained by  performing the change of variables \eqref{ggrungoGringo} in the first equation of \eqref{hydroequez}. 
Considering that there is a gauge freedom which can be used to change the value of one free function without altering the physical state of the system \citep{Comer_1993}, we are left again with $\mathfrak{D}=2\times 3-1=5$.

\section{Slow limit of the non-interacting scalar field}

We explicitly discuss the slow limit procedure, by using the complex scalar field as a guiding example. In its simplicity, the Klein-Gordon model contains most of the field theory features we are interested in. 

\subsection{The slow limit removes degrees of freedom}

The slow limit is a necessary element to extract (in ergodic systems of interacting identical particles with no broken symmetries) the universal Navier-Stokes equations. The main reason is that this limit removes all the fast degrees of freedom. Let us see how this mechanism of suppression of degrees of freedom works for the non-interacting scalar field. We anticipate that this slow limit can not be Navier-Stokes, since the existence of an ``order parameter'' prevents the emergence of a universal behaviour \cite{ruffini69}. 
In fact, it is well known that the slow limit of the Klein-Gordon equation is the Schr\"{o}dinger equation: following the standard textbook approach (e.g. \citep{Zee2003}), it is possible to define  
\begin{equation}\label{wave}
\varphi =  e^{-imt} \psi \, / \, \sqrt{2m} 
\end{equation}
that, when used into \eqref{klinGordon}, allows to obtain
\begin{equation}\label{KKgor}
- \, \dfrac{1}{2m} \partial_t^2 \psi +i  \partial_t \psi = - \, \dfrac{1}{2m} \partial_j \partial^j \psi \, .
\end{equation}
To recover the slow limit we have to recall that $1/m$ sets an intrinsic timescale: demanding $\psi$ to be  a slow degree of freedom amounts to require that
\begin{equation}\label{chain}
|\partial_t^2 \psi| \ll m|\partial_t \psi| \, .
\end{equation}
In this limit, we obtain, as expected,
\begin{equation}\label{shrodo}
i  \partial_t \psi = - \, \dfrac{1}{2m} \partial_j \partial^j \psi \, .
\end{equation}
%
Contrarily to the original Klein-Gordon equation, equation \eqref{shrodo} is of the first-order in time, meaning that we are left only with $\mathfrak{D}=2$. To see what has happened to the remaining 2 degrees of freedom, we can take the small-momentum limit ($p_j \longrightarrow 0$) of the generic solution \eqref{soluzionediklein} and cast it into the form
\begin{equation}\label{2wave}
\varphi =\dfrac{1}{\sqrt{2m}} \big( e^{-imt} \psi_+ + e^{imt} \psi_- \big) \, ,
\end{equation}
where we have introduced the two functions
\begin{equation}\label{twowaves}
\begin{split}
& \psi_+(x) =  \int  a_p \exp\bigg(i p_j x^j-i  \dfrac{p_j p^j}{2m} t \bigg) \, \dfrac{d^3p}{(2\pi)^3} \\
& \psi_-(x) = \int  \bar{b}_p \exp\bigg(-i p_j x^j + i  \dfrac{p_j p^j}{2m} t \bigg) \, \dfrac{d^3p}{(2\pi)^3}
\, . 
\end{split}
\end{equation}
Direct comparison of \eqref{wave} with \eqref{2wave} gives
$\psi = \psi_+ + e^{i2mt} \psi_-$.
Since $e^{i2mt}$ oscillates fast, while $\psi_+$ and $\psi_-$ are slowly evolving, the only way for $\psi$ to fulfill \eqref{chain} is that
\begin{equation}\label{psim}
\psi_- \, = \, 0    \qquad 
\Longrightarrow
\qquad \psi = \psi_+  \, .
\end{equation}
This condition is the reason for the halving of $\mathfrak{D}$ in the slow limit. 
In fact, from \eqref{twowaves} it is evident that $\psi_+$ and $\psi_-$ are two independent functions, where $\psi_+$ is governed by \eqref{shrodo}, while $\psi_-$ is governed by the time-reversed Schr\"{o}dinger equation. 
Thus, the couple $(\psi_+,\psi_-)$ constitutes a convenient representation of the $\mathfrak{D}=4$ degrees of freedom of the theory for small momenta. Only when the slow-limit is taken we have
\begin{equation}
(\psi_+,\psi_-) \, \longrightarrow \, (\psi_+,0)
\end{equation}
and the second-order (in time) Klein-Gordon equation boils down to a first-order one. 
To understand the physical implications of \eqref{psim}, consider the total four-momentum $\mathcal{P}^\nu $ and the $U(1)$ charge $Q$ of the system:
\begin{equation}\label{hinomkb}
\begin{split}
& \mathcal{P}^\nu :=\! \int T^{0\nu} \, d^3 x  
= \! \int p^\nu \big(\bar{a}_p a_p + \bar{b}_p b_p  \big)\dfrac{d^3p}{(2\pi)^3} 
\\
& Q :=\! \int n^{0} \, d^3 x 
 =\! \int \! \big(\bar{a}_p a_p - \bar{b}_p b_p  \big)\dfrac{d^3p}{(2\pi)^3} \, ,
 \end{split}
\end{equation}
where $\bar{a}_p a_p$ and $\bar{b}_p b_p$ can be interpreted as the average number of particles and antiparticles with four-momentum $p^\nu$. Hence, by imposing equation \eqref{psim}, which implies $\bar{b}_p b_p=0$, we are excluding antiparticle excitations from the model \citep{Heyen2020}. 

In conclusion, we have shown that the slow limit of a field theory has, in general, less degrees of freedom than its fast counterpart. As a result, when we impose a condition of slow evolution, we typically constrain the initial conditions to belong to a particular subset (of measure zero) of the full state-space. 

\subsection{Dispersion relations}\label{Latechnicadeimodi}
 
Thanks to the minimal example provided by the Klein-Gordon model, we have seen that the slow-limit lowers the order of temporal derivatives of the dynamical equations. 
It is natural to ask whether there is a more systematic technique to study the effect of the slow limit also on other systems. 
A way of doing this is via  the spectral analysis of the modes of the linearised model around a homogeneous stationary configuration. 
What follows is a discussion of the results of the previous subsection from this, more general, point of view.

\begin{figure}
\begin{center}
	\includegraphics[width=0.49\textwidth]{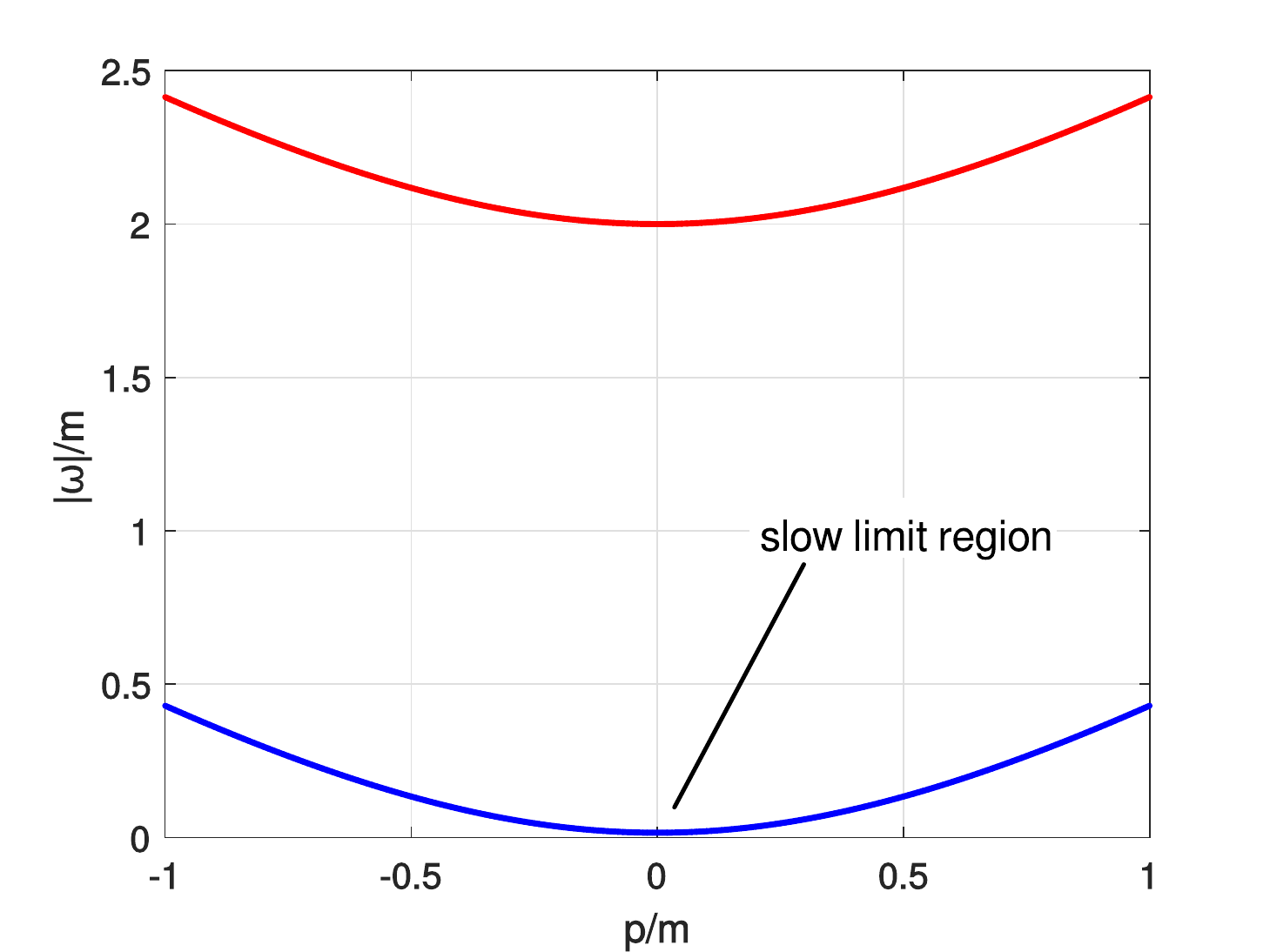}
	\caption{Absolute value of the dispersion relations of the complex Klein-Gordon field $\psi$, as given in equation \eqref{odkmc}. The blue curve refers to the gapless mode $\omega_+$, the red line is the gapped mode $\omega_-$. }
	\label{fig:gapped}
	\end{center}
\end{figure}

We consider solutions of the full Klein-Gordon equation \eqref{KKgor} of the kind
\begin{equation}
\psi= \psi_+^{(0)} + \delta \psi \, e^{i(p_j x^j-\omega t)} \, ,
\end{equation} 
where $\psi_+^{(0)}$ and $\delta \psi$ are constants; the pre-factor $\delta \psi$ is infinitesimal. Since the hydrodynamic equation is already linear, the factor $\exp(ip_j x^j-i\omega t)$ must be an exact solution of \eqref{KKgor}.   
For every wave-vector $p^j$, there are two oscillation modes of  $\psi$, with frequencies
\begin{equation}\label{odkmc}
\omega_{\pm} = -m \pm \sqrt{m^2 + p^j p_j} \, .
\end{equation}
The absolute values of these dispersion relations are presented in Figure \ref{fig:gapped}. In the small wave-vector limit, equation \eqref{odkmc} becomes
\begin{equation}
\omega_+ \approx \dfrac{p^j p_j}{2m}  \quad  \quad \quad    
\omega_- \approx -2m- \, \dfrac{p^j p_j}{2m} \, .
\end{equation}
The  mode evolving with $\omega_+$  contributes to $\psi_+$, while the  mode evolving with $\omega_-$ contributes to $\psi_-$. The small-momenta behaviour of the two modes is
\begin{equation}
\lim_{p \rightarrow 0} \left\lvert \, \omega_+ \right\rvert =0
\quad \quad \quad 
\lim_{p \rightarrow 0} \left\lvert \, \omega_- \right\rvert =2m \neq 0 \, ,
\end{equation}
and  we will refer to $\omega_+$ as the \emph{gapless} mode, and  to $\omega_-$ as the \emph{gapped} mode, with gap $2 m$.

Figure \ref{fig:gapped} gives an intuition of what happens in the slow limit $\omega \rightarrow 0$: the full solution must be a superposition of modes which are in the proximity of the x-axis when the slow limit is taken. Since the gapped dispersion relation does not touch the x-axis, in the limit of small $\omega$ the full solution must be a superposition of only the $\omega_+$ modes, so that equation \eqref{psim} is enforced  automatically. Imposing that the final solution can be fully described with only one dispersion relation ($\omega_+$) implies that we are replacing a second-order differential equation with a first-order one, and this is exactly what happens in the slow limit.

\subsection{Hydro-modes and Thermo-modes}\label{HydroThermo}

It is instructive to give a physical interpretation to the presence and the role of both gapped and gapless modes. We will limit ourselves to systems that do not exhibit  the so-called \emph{gapped momentum modes}, see e.g. \citet{BAGGIOLI20201}.

First, let us focus on the modes with gapless dispersion relation $\omega_+$. As they contribute only to $\psi_+$, we can conclude that they describe ordinary density fluctuations in the medium. They are, essentially, the analogue of sound-waves: we  call them \textit{Hydro-modes}, because they are (in this context) the common  hydrodynamic modes expected in practically every fluid. 
In our minimal example the dispersion relation $\omega_+$ is not linear (i.e. we do not have a phonon-like dispersion relation) simply because the system is free: by adding a term of the kind $\bar{\varphi}^2\varphi^2$ in the action \eqref{action}, we would recover the expected linear phonon-like behaviour of $\omega_+$ for small momenta. 

On the other hand, the modes with gapped dispersion relation $\omega_-$ are present whenever the antiparticle fraction is non-zero. 
Hence, contrarily to the Hydro-modes, these modes do not simply describe the evolution  of an inhomogeneity (i.e. a deviation from the homogeneous state). They give rise, instead, to a non-stationary behaviour that survives even in the absence of spatial gradients: this is the physical meaning of the gap. 
We can gain further intuition if we imagine to switch on a small interaction term in the action. Now, particle-antiparticle reactions are allowed and the antiparticle fraction is rapidly converted into thermal energy. Therefore, states with $\psi_- \neq 0$ are out of equilibrium with respect to pair production/annihilation processes. We can conclude that the gapped modes arise from the possibility to have homogeneous states that are, however, non-equilibrium states (a situation that is typical when it comes to model chemically reacting substances). Since the theory which deals with such reacting systems is non-equilibrium thermodynamics \citep{prigoginebook}, we may call these modes \textit{Thermo-modes}. 

In the next section we will show that the division of the modes of the full (non-slow) theory into Hydro-modes and Thermo-modes still holds also for more general systems of interacting particles and is key to understand the emergence of hydrodynamics as the slow limit of more fundamental theories. 
We anticipate that, in a generic dissipative fluid,  in principle we may  deal with an infinite number of degrees of freedom, that generate an infinite number of dispersion relations \citep{Denicol_Relaxation_2011}. 
However, almost all these dispersion relations are gapped (the associated Thermo-modes describe the tendency of the fluid elements to relax towards local thermodynamic equilibrium). The only modes that survive in the slow limit are the few gapless ones, which describe the transport of conserved quantities \cite{Glorioso2018} and are (almost \cite{jain_kovtun_2020arXiv}) universally described by the Navier-Stokes equations (another reason we call them Hydro-modes). This mechanism will be explored in detail in subsection \ref{degmuiono}, starting from a kinetic description. 
A similar argument is at the origin of the \textit{hydrodynamization} mechanism in heavy ion collisions \citep{FlorkowskiReview2018}. 

\subsection{Lorentz boosts can generate spurious gapped modes}\label{spuriii}

We briefly discuss why the slow limit can be problematic in relativity. 
So far, we worked in the reference frame in which the slow limit was taken. As long as the description is limited to this frame, there is no fundamental difference between the Newtonian and the relativistic descriptions of spacetime. 
Problems with the slow limit in relativity, however, appear when we decide to work in a different reference frame from the one in which the limit itself is taken. To see exactly what can happen, we consider the minimal example of the Schr\"{o}dinger limit of the Klein-Gordon model.

Let us assume that an observer (say Alice, moving with four-velocity $u_A^\nu$) has prepared a condensate that is at rest in her laboratory, while Bob (who is moving with four-velocity $u_B^\nu \neq u_A^\nu$), wants to describe the evolution of this same condensate in his own reference frame. 
Bob may start from the assumption that the condensate is in the slow limit from Alice's point of view and, therefore, argue that equation \eqref{shrodo} should hold in her reference frame. 
Recalling the definition \eqref{wave}, Bob will conclude that $\varphi$ obeys the  equation (formulated on the basis of the principle of covariance)
\begin{equation}\label{ghijrmf}
i u_A^\nu \partial_\nu \big( \varphi \, e^{-im u_A^\sigma x_\sigma} \big) 
= 
- \big( g^{\nu \rho} + u_A^\nu u_A^\rho \big) \dfrac{ \partial_\nu \partial_\rho}{2m}
\big( \varphi \, e^{-im u_A^\sigma x_\sigma} \big) \, .
\end{equation}
In this way, Bob is making sure that any solution he may find will correspond automatically to an exact solution of the  Schr\"{o}dinger equation in Alice's reference frame. 

There is, however, a complication. In fact, in every reference frame, apart from the one of Alice, equation \eqref{ghijrmf} is of the second order in time, implying the paradoxical conclusion that Lorentz boosts double the number of degrees of freedom of the Schr\"{o}dinger field. 
In fact, in the reference frame of Bob, there are two dispersion relations, instead of just one. On the other hand, if $u_B^\nu$ approaches $u_A^\nu$, we  recover the usual Schr\"{o}dinger equation, which predicts only one dispersion relation. What happens to the second one? 

To answer this question, we look for solutions of \eqref{ghijrmf} which are homogeneous in Bob's reference frame,
\begin{equation}\label{vdgub}
\varphi \propto e^{i \, (m+\omega) \, u_B^\sigma x_\sigma} \, ,
\end{equation}
as they  allow for an immediate estimate of the possible gaps of the two dispersion relations observed by Bob. Plugging \eqref{vdgub} into \eqref{ghijrmf} we obtain
\begin{equation}\label{gbkmvg}
(\gamma^2-1) \, \omega^2 + 2m (\gamma^2-\gamma-1) \, \omega + m^2(\gamma-1)^2 =0 \, ,
\end{equation}
where $\gamma=-u_A^\nu u_{B\nu}$ is the usual Lorentz factor. The two solutions of  \eqref{gbkmvg} are
\begin{equation}\label{mokhioj}
\dfrac{\omega_{\pm}}{m} = \dfrac{1 \pm \sqrt{1-2w^2}}{\gamma w^2} -1 \, ,
\end{equation}
where $w$ is the relative speed between the two observers, namely $\gamma^{2} =1/( 1-w^2)$.
For small relative speed the two oscillation modes in \eqref{mokhioj} read 
\begin{equation}\label{jkub}
\omega_+ \approx \dfrac{2m}{w^2}   \quad \quad \quad \quad    \omega_- \approx \dfrac{m w^4}{8} \, .
\end{equation}  
Hence, in the limit $w \rightarrow 0$ of comoving observers, we have that $\omega_-$ becomes the homogeneous zero-frequency mode of the Schr\"{o}dinger equation. On the other hand, $\omega_+$ diverges. To understand this peculiar behaviour of $\omega_+$ we ask how its associated mode looks like in the reference frame of Alice; plugging \eqref{jkub} into \eqref{vdgub} and recalling \eqref{wave}, we obtain
\begin{equation}\label{ilproblematicoAAA}
\psi \propto \exp \bigg[i \dfrac{2m}{u_B^k u_{Bk}} (u_B^j x_j-t)  \bigg] \, .
\end{equation}
It can be directly verified that this is indeed a solution of \eqref{shrodo}. However, we also see that its associated wave-vector is
\begin{equation}
p^j = \dfrac{2mu_B^j}{u_B^k u_{Bk}} \, ,
\end{equation}
which diverges in the limit of small $u_B^j$. Therefore, we can conclude that this mode does not respect the assumption of slow evolution in the Alice's frame. This, in turn, implies that the Schr\"{o}dinger equation is not applicable along this mode, which should, then, be discarded as an unphysical solution.

To summarize, we have shown that in the reference frame of Bob there are two dispersion relations $\omega_+$ and $\omega_-$ (instead of just one as in Alice's reference frame), where $\omega_+$ is associated with an unphysical gapped mode, whose gap diverges when $u_B^\nu \longrightarrow u_A^\nu$. 
This is the prototype of what we will refer to as \emph{spurious} mode, namely an additional (and unphysical) gapped mode that may emerge as a consequence of the straightforward uplift of a legit Newtonian model to a covariant one, as done in \eqref{ghijrmf}.

\subsection{A more general discussion of the spurious modes} 

The example analysed in the previous subsection is a particular case of a more general issue, whose aspects are qualitatively summarised in the following points:

\begin{enumerate}
\item[i-] When the slow limit of a covariant theory is taken in a given reference frame $A$, space and time are not treated on equal footing anymore: the slow limit typically involves some truncation in the time-derivative expansion, so that the order in time of the final equations is typically lower than the order in space.
\item[ii-] When we boost to a reference frame $B \neq A$, the derivatives in space become linear combinations of both derivatives in space and time,
\begin{equation}
\dfrac{\partial}{\partial x} = \gamma \dfrac{\partial}{\partial x'} -\gamma w \dfrac{\partial}{\partial t'} \, ,
\end{equation}
increasing the order in time of the equations. 
This gives rise to gapped dispersion relations $\omega_B$ which do not exist in the Newtonian theory \citep{GavassinoLyapunov_2020}.
\item[iii-] As $B \longrightarrow A$, the dispersion relations $\omega_B$ cannot change smoothly, because they do not exist for $B=A$. The limit  
\begin{equation}\label{mdokcploki}
\lim_{B \rightarrow A} \omega_B
\end{equation}
usually results in an infinity, showing that the corresponding mode is unphysical.
\item[iv-] Therefore, if we want any solution of the boosted system of equations to be physical, we must first make sure that it does not contain any contribution coming from the modes $\omega_B$, at least to the level of precision we are interested in.  
\end{enumerate}

The last point is what can make the slow limit problematic in relativity. In fact, depending on the sign of $\text{Im}(\omega_B)$, achieving the condition in (iv) may be easy, difficult or even impossible for some systems.

When $\text{Im}(\omega_B) <0 $,
then the spurious unphysical modes are naturally damped; whatever the initial condition, the solution converges to a physical one and the predictions of the model become reliable after a time of a few $1/|\text{Im}(\omega_B)|$. This is the ``easy'' case.

If $\text{Im}(\omega_B) = 0 $,
as in the case of the Schr\"{o}dinger equation (for $2w^2 \leq 1$), these modes oscillate with constant amplitude: if the gapped modes are negligible at the beginning of the evolution, they will not become important later. Hence, if we want to find solutions which are physically acceptable, we have to enforce the (approximate) absence of the spurious modes in the initial conditions.  This tuning of the initial condition may be complicated to achieve, but it does not come unexpected. In fact, when Bob makes the assumption that the fluid is slow in the reference frame of Alice, it becomes immediately clear that he is restricting his attention to those solutions that are compatible with this constraint.

Finally, it may happen that the unphysical modes grow exponentially,
$ \text{Im}(\omega_B) >0 $. 
Even if we fine-tune the initial conditions in such a way to approximately remove the spurious modes, these modes will appear after a time-scale $1/\text{Im}(\omega_B)$. 
The only way to avoid this explosion would be to select some initial conditions in which the unphysical modes have been  removed exactly, making the theory impossible to use in numerical simulations, where one is forced to deal with a finite precision.

In the next section we verify that, unfortunately, the spurious modes of the  Navier-Stokes-Fourier approach display an explosive behaviour ($ \text{Im}(\omega_B) >0 $).

\section{The role of the slow limit: the diffusion equation}\label{RadiatttT}

In this section, we consider the case of  photon diffusion in a medium to see in more detail how the Navier-Stokes-Fourier approach\footnote{
We recall that by ``Navier-Stokes-Fourier approach'' we mean the assumption that the dissipative fluxes, such as heat flux and viscous stresses, are proportional to the spatial gradients of a corresponding perfect-fluid field, such as the temperature or the fluid velocity. 
}
emerges in the slow limit of a relativistic dissipative system. For simplicity, we work under the assumption of a perfectly rigid medium in a flat spacetime. 
Our purpose is not to provide a framework for the realistic description of the photon-matter system, but rather to illustrate the role of the slow limit with an example. More complete discussions of radiation hydrodynamics can be found in \citet{AnileRadiazion1992,UdeyIsrael1982,Farris2008,Sadowski2013,GavassinoRadiazione}. 

\subsection{Minimal model for photon diffusion in a material}
  
Consider a photon gas diffusing in a rigid material. Since the total number of photons is not a conserved charge, the only relevant conserved current $n^\nu$ is the baryon current of the material (we assume that no additional chemical-composition degrees of freedom are needed). This system can be described within the assumptions of subsection \ref{iPPhi}: despite there are two chemical species (matter and photons), the photon transport is just a form of heat conduction, as no transport of any Noether charge is involved \citep{GavassinoRadiazione,GavassinoTermometri}. 

We define the material rest-frame density and four-velocity as, respectively,
\begin{equation}
n=\sqrt{-n^\rho n_\rho}    \quad \quad   \quad \quad   u^\nu = {n^\nu}/{n} \, , 
\end{equation}
and assume the material to be rigid (in the Born sense), non-rotating and non-accelerating, namely  
\begin{equation}\label{eofmvc}
\nabla_\beta \,  u^\nu = 0 \, .
\end{equation}
This condition implies that we can introduce a conserved energy current $J^\nu $ as
\begin{equation}\label{fermati}
J^\nu =- T^{\nu \rho}u_\rho 
\qquad  \Rightarrow  \qquad 
\nabla_\nu J^\nu =0 \, .
\end{equation}
Its associated conserved charge
\begin{equation}
U = -\int_\Sigma J^\nu \, \,  d\Sigma_\nu
\end{equation}
is the total energy of the system, as measured in the global inertial frame defined by $u^\nu$, no matter which Cauchy hypersurface $\Sigma$ is chosen. In the following, we will focus on giving a constitutive relation only to $J^\nu$, ignoring the remaining parts of the stress-energy tensor.

Finally, to further simplify the analysis, we also require that the rest-frame density $n$ is homogeneous, $\nabla_\rho \, n = 0$.
Combining this condition with \eqref{eofmvc}, we get the particle conservation $\nabla_\nu n^\nu=0$ as an identity. Hence, no matter which fields $\varphi_i$ we choose, the constitutive relation for $n^\nu$ will always be $n^\nu = \text{const}$.

\subsection{Photon diffusion: constitutive relations and hydrodynamic equations}
\label{JonasCurioso}
 
Following \citet{Weinberg1971}, we assume that the mean free path and time of the particles comprising the material is infinitely shorter than the one of the photons. Therefore, if we assume small deviations from equilibrium but \textit{not} slow evolution, then the simplest effective field theory that is consistent with radiation kinetic theory \citep{mihalas_book} is the $\text{M}_1$ closure scheme \citep{Sadowski2013}.

In view of the assumptions made in the previous subsection, a minimal $\text{M}_1$ model for the diffusion of photons in a rigid material can be built by using only two fields,
$(\, \varphi_i \,)  =  (\,\Theta\, , \, q^\sigma\,) \,$.
The field $q^\sigma$ can be interpreted as the heat flux (as measured in the frame of $u^\sigma$) and satisfies the constraint $q^\sigma u_\sigma =0$.
The scalar $\Theta$ is the temperature of the material, which is in local thermodynamic equilibrium, due to the infinitesimal mean free time of its constituents. If we require that the radiation gas is sufficiently close to local thermodynamic equilibrium, with the same temperature of the material $\Theta$ (see Section \ref{TwoTemper} for the case with different temperatures), then we may assume that the internal energy density
\begin{equation}
\mathcal{U} = T^{\nu \rho}u_\nu u_\rho = -J^\nu u_\nu
\end{equation}   
is given by the equilibrium matter+radiation equation of state. We can thus postulate the  constitutive relation \cite{Eckart40}  
\begin{equation}\label{JnuJ}
J^\nu = \mathcal{U} \, u^\nu + q^\nu   \quad \quad \quad \quad
\mathcal{U}=\mathcal{U}(\Theta) \, ,
\end{equation}
while the constitutive relation for the entropy current will not be necessary.

Given the relevant constitutive relations, we now have to specify the hydrodynamic equations, which can be derived from kinetic theory. For simplicity, let us assume that photons in the material undergo only absorption/emission processes (i.e. scattering is negligible). Thus, if we work in an inertial frame, the photon one-particle distribution function $f$ obeys a  transport equation of the kind \citep{AndersonWitting1974,cercignani_book,UdeyIsrael1982}
\begin{equation}\label{pafnu}
p^\nu \partial_\nu f = -p^\nu u_\nu \chi \, \big(f_{\text{eq}}-f \big) \, ,
\end{equation}
where $\chi>0$ is the absorption opacity (that we assume to be independent from the photon energy), while $f_{\text{eq}}$ is the usual equilibrium Bose-Einstein distribution function, 
\begin{equation}\label{BoseEinstein}
f_{\text{eq}} =\dfrac{2}{(2\pi)^3} \dfrac{1}{e^{-p^\nu u_\nu/\Theta}-1} \, .
\end{equation}
The second moment of the distribution function $f$ is the radiation stress-energy tensor \citep{Groot1980RelativisticKT},
\begin{equation}\label{ventinove}
R^{\nu \rho} = \int f \, p^\nu p^\rho \, \dfrac{d^3 p}{p^0} \, ,
\end{equation}
that must satisfy an equation analogous to \eqref{pafnu}, namely
\begin{equation}\label{RaZZAZA}
\partial_\nu R^{\nu \rho} = -\chi \, u_\nu (R^{\nu \rho}_{\text{eq}}-R^{\nu \rho}) \, ,
\end{equation}
where $R^{\nu \rho}_{\text{eq}}$ is just $R^{\nu \rho}$  but computed  with $f_{\text{eq}}$ in place of $f$. Working in the global inertial frame defined by $u^\nu$, and recalling that the material is in local thermodynamic equilibrium, we find
\begin{equation}\label{seccondo}
R^{0 j}_{\text{eq}} =0   \quad \quad  \quad \quad    R^{0j}=q^j \, .
\end{equation}  
Furthermore, close to local thermodynamic equilibrium, the  $\text{M}_1$  closure scheme imposes \citep{GavassinoRadiazione}
\begin{equation}\label{primmo}
R^{kj}=\dfrac{1}{3} a_R  \Theta^4 \, \eta^{kj} \, ,
\end{equation}
where $a_R$ is the radiation constant. Using \eqref{seccondo} and \eqref{primmo} into the $j$-th component of \eqref{RaZZAZA} gives 
\begin{equation}
\partial_t q_j + \dfrac{4}{3} a_R \Theta^3 \partial_j \Theta = -\chi q_j \, .
\end{equation} 
Recalling that also equation \eqref{fermati} must be valid, we end up with the system of hydrodynamic equations
\begin{equation}\label{systemmum}
 c_v \partial_t \Theta + \partial_j q^j=0 
 \qquad \qquad
 \tau \partial_t q_j + q_j = -\kappa \partial_j \Theta \, ,
\end{equation}
where, in complete agreement with \citet{Weinberg1971}, the coefficients are defined as
\begin{equation}\label{quattordici}
c_v = \dfrac{d \mathcal{U}}{d \Theta}  \quad \quad    \quad
\tau = \dfrac{1}{\chi} \quad \quad    \quad
\kappa = \dfrac{4 a_R \Theta^3}{3\chi} \,  ,
\end{equation}
which can be interpreted as the heat capacity per unit volume (at constant volume) of the  matter+radiation system, the heat relaxation-time and the heat conductivity. The system \eqref{systemmum} is comprised of 4 first-order equations for the  4 functions $(\Theta,q^j)$, implying that $\mathfrak{D}=4$. The second equation of \eqref{systemmum} is known as the Maxwell-Cattaneo equation \cite{Jou_Extended}.

\subsection{Photon diffusion: linear analysis}

Let us perform, for small deviations from equilibrium, a linear analysis of the effective field model derived in the previous subsection. Taking the divergence of the second equation in \eqref{systemmum}, using the first one and linearising the result, we obtain the telegraph-type equation \cite{rezzolla_book}
\begin{equation}\label{rtkmovklll}
\tau \,  \partial_t^2  \Theta + \partial_t \Theta = \mathcal{D} \,  \partial_j \partial^j \Theta \, ,
\end{equation} 
where we have introduced the photon-diffusion coefficient $\mathcal{D} = \kappa/c_v$. The similarities between this equation and \eqref{KKgor} are evident. 
In the slow limit, formally given by 
$ \tau |\partial_t^2 \Theta| \ll |\partial_t \Theta| $, 
the Cattaneo equation boils down to the diffusion equation\footnote{
The coefficient $\mathcal{D}=\kappa /c_v$ differs from the standard thermal diffusivity coefficient $\kappa/c_p$ \citep{landau6} because we are working under the simplifying assumption of a perfectly rigid material (in particular, $\partial_t n=0$). Indeed, the rigidity assumption is usually justified in solids, where $c_p \approx c_v $ \citep{landau5}.
}
\begin{equation}\label{demifr}
\partial_t \Theta = \mathcal{D} \,  \partial_j \partial^j \Theta \, .
\end{equation} 
This fact allows us,  in analogy with subsection \ref{HydroThermo}, to interpret the gapless modes of the theory as the Hydro-modes of the photon gas. In fact, equation \eqref{demifr} describes how the photons tend to diffuse in the material, by following a random walk, until $\Theta$ becomes homogeneous. 

To probe the possible presence of gapped modes we only need to take the homogeneous limit of the system \eqref{systemmum}, obtaining
\begin{equation}
\partial_t \,  \Theta = 0   \quad \quad \quad   \tau \,  \partial_t \,  q_j =- q_j  \, .
\end{equation}
We can, therefore, conclude that all the homogeneous solutions of the theory are
\begin{equation}\label{Igappedcrudeli}
\Theta = \text{const}   \quad \quad \quad  q_j(t)=q_j(0) \, e^{-t/\tau}       
\end{equation}
which are the homogeneous limit of three independent gapped modes, one for each component $q_j$.


\subsection{Which degrees of freedom are suppressed in the slow limit?}\label{degmuiono}

It is interesting to see  what happens when we make the transition to the slow limit directly in the context of kinetic theory. We will use our toy-model for photon diffusion, but analogous arguments are also used to derive the Navier-Stokes-Fourier approach from the kinetic theory of ideal gases \citep{huang_book,cercignani_book}. 
Let us introduce the function $ h =f-f_{\text{eq}} $, representing the deviation of the one-particle distribution $f$ from the equilibrium distribution \eqref{BoseEinstein}.
Thanks to  the second definition in \eqref{quattordici}, the kinetic equation \eqref{pafnu} becomes
\begin{equation}\label{inhomoGG}
\dfrac{\tau}{p^0} p^\nu \partial_\nu h + h = 
- \, \dfrac{\tau}{p^0} p^\nu \partial_\nu f_{\text{eq}} \, .
\end{equation}
From equation \eqref{BoseEinstein}, we know that $f_{\text{eq}}(x,p)=f_{\text{eq}}(\Theta(x),p)$, which tells us that the evolution of $f_{\text{eq}}$ is completely determined by the Cattaneo equation. Hence, equation \eqref{inhomoGG} can be seen as a non-homogeneous partial differential equation for $h$, with the right-hand side playing the role of the source term. 
Its solution, then, reads
\begin{equation}\label{hoftitty}
h(t) =h(0)e^{-t/\tau} - \dfrac{\tau}{p^0} \int_0^{t/\tau} \bigg( p^\nu \partial_\nu f_{\text{eq}} \bigg)\bigg|_{t-\tau \xi} e^{-\xi} \, d\xi \, ,
\end{equation}
where all the quantities are computed along the geodesic path in phase space \citep{MTW_book} with parameter $t$  
\begin{equation}\label{geoddetica}
\bigg(t, \, x^j+\dfrac{p^j}{p^0}t , \, p^j  \bigg) \, .
\end{equation}
Now, we can finally explain what happens when we take the slow limit.
Since equation \eqref{inhomoGG} is a dynamical equation for $h$, in the solution \eqref{hoftitty} we are free to choose the initial conditions $h(0)$ at will, provided that they respect the assumptions underlying the Cattaneo equation. 
This is why the heat flux
\begin{equation}\label{maqj}
q^j =\int h \, p^j \, d^3 p \, ,
\end{equation}
see equations \eqref{ventinove} and \eqref{seccondo}, is a degree of freedom of \eqref{systemmum}: it represents the freedom of preparing the photons with arbitrary initial momenta. However, after a collision time-scale $\tau$, all these momenta are randomised and we lose information about the original distribution $h(0)$. This is reflected in the fact that, for $t \gg \tau$, equation \eqref{hoftitty} can be approximated as
\begin{equation}\label{hoftitty2}
h(t) \approx - \dfrac{\tau}{p^0} \int_0^{+\infty} \bigg( p^\nu \partial_\nu f_{\text{eq}} \bigg)\bigg|_{t-\tau \xi} e^{-\xi} \, d\xi \, .
\end{equation}
This loss of information about the initial non-equilibrium part of the distribution function (a process that can occur also in the homogeneous limit) is captured, at the level of equations \eqref{systemmum}, by solutions of the form \eqref{Igappedcrudeli}. These be can interpreted as transient phases in which an initial flux of photons is rapidly damped by the collision processes which randomize the photons' momenta. This is the dynamics expected for a Thermo-mode, as discussed in subsection \ref{HydroThermo}.

After the aforementioned relaxation process has occurred, the only non-equilibrium feature that survives are large-scale inhomogeneities  \cite{Glorioso2018}, which can now be safely assumed as slowly-evolving. This allows us to impose
\begin{equation}
\big( \, p^\nu \partial_\nu f_{\text{eq}} \, \big)\big|_{t-\tau \, \xi} 
\, \approx  \, 
\big( \, p^\nu \partial_\nu f_{\text{eq}} \, \big)\big|_{t}  
\quad  \text{for} \quad  \xi \lesssim 1
\end{equation}
and we finally obtain
\begin{equation}\label{hhhh}
h \approx - \dfrac{\tau}{p^0}  p^\nu \partial_\nu f_{\text{eq}} \, .
\end{equation}
This equation is similar to \eqref{inhomoGG}, with the difference that we have removed the first term on the left-hand side. This, however, changes completely the character of the equation, as it converts it from a dynamical equation for $h$ into a constraint: \eqref{hhhh} is a relation where the value of $h$ is expressed in terms of the gradient of $\Theta$, which is the field that defines $f_{\text{eq}}$. 
In fact, inserting \eqref{hhhh} into \eqref{maqj}, it is easy to show that
\begin{equation}\label{FFouRRieRRzxub}
q_j = -\kappa \, \partial_j \, \Theta \, ,
\end{equation}  
which is the usual Fourier law. Plugging it into the first equation of \eqref{systemmum} we recover \eqref{demifr}.

In conclusion, we have shown that the slow limit has the effect of downgrading the heat flux $q_j$ from a degree of freedom to a quantity whose value is completely fixed in terms of another field of the model, here $\Theta$. 
This means that, when working  directly in the slow limit, we are not free to set $h(0)$ arbitrarily. 

In a fully non-linear regime (but under the assumption of small spatial gradients), the tendency of dissipative fluids to lose degrees of freedom, by transforming dynamical equations into phenomenological constraints, has been rigorously proved by \citet{LindblomRelaxation1996}, who called this process \textit{Relaxation Effect}. He showed that, after an initial transient (which is nothing but the non-linear analogue of the initial fast decay of the Thermo-modes), dissipative fluids asymptotically relax to physical states that are essentially indistinguishable from Navier-Stokes fluids.

\subsection{Boost-generated spurious modes}\label{ilsuccodelproblema}

Analogously to the Schr\"{o}dinger equation, also the heat equation \eqref{demifr} produces a spurious gapped mode when boosted \citep{Kost2000}. Below, we study this spurious mode following the same reasoning as in subsection \ref{spuriii}. 

Assume that the material is at rest (and the system is slowly evolving) in the Alice frame defined by $u_A^\nu=u^\nu$. Again, Bob, moving with $u_B^\nu \neq u_A^\nu$, models the evolution of the photon gas in his own reference frame by assuming  that   \eqref{demifr} holds in Alice's frame, and promoting it to the covariant form
\begin{equation}\label{zentratre}
u_A^\nu \partial_\nu \Theta = \mathcal{D} (g^{\nu \rho} + u_A^\nu u_A^\rho) \partial_\nu \partial_\rho \Theta \, .
\end{equation}
This equation is of the second order in time in any reference frame, apart from the one of Alice (in the frame of Bob there will be two dispersion relations, instead of just one). To study the possible presence of gaps, we look for solutions which are homogeneous in Bob's reference frame,
\begin{equation}
\Theta \propto e^{i \,\omega  \,u_B^\sigma  \,x_\sigma} \, .
\end{equation} 
This converts equation \eqref{zentratre} into the algebraic equation
\begin{equation}
i \mathcal{D} \gamma w^2 \omega^2 + \omega =0 \, ,
\end{equation}
whose two solutions are
\begin{equation}
\omega_A =0   \qquad \qquad \quad   
\omega_B = \dfrac{i}{\mathcal{D} \gamma w^2} \, .
\end{equation}
We see that the gapped dispersion relation has exactly the same pathological character discussed in the previous section: as $u_B^\nu \rightarrow u_A^\nu$ the gap diverges and it can be verified that its associated mode is unphysical as it is strongly acausal \cite{GavassinoLyapunov_2020}. 
Furthermore, we have that $\text{Im}(\omega_B)>0$, making the diffusion equation impossible to use (in Bob's reference frame) with any finite precision method. 

The existence of the unphysical gapped mode $\omega_B$, which makes the initial value problem ill-posed \citep{Kost2000}, is at the origin of the generic instability problem of first-order theories described by \citet{Hiscock_Insatibility_first_order}. In fact, the theories of \citet{Eckart40} and \citet{landau6} predict spurious modes of the kind exemplified above. In view of the present discussion, this is not surprising, since they arise from the straightforward application of the Navier-Stokes-Fourier approach in the fluid's rest-frame. 

\subsection{Is the Navier-Stokes-Fourier approach wrong?}

Before moving on with the general discussion, let us briefly comment on a natural question that may arise at this point: is equation \eqref{zentratre} ``wrong''? 

Since the Navier-Stokes-Fourier approach describes fluids which turn out to be acausal and unstable, it is possible to interpret this fact as a sign that this approach should be discarded in relativity \citep{Hiscock_Insatibility_first_order,rezzolla_book}. 
On the other hand, \citet{Kost2000} pointed out that, for realistic physical states (and therefore in the absence of spurious modes), the superluminal propagation of signals happens only on small time-scales and length-scales. 
Hence, given that the Navier-Stokes description is an approximation for the large-scale and slow behaviour of the fluid (namely, it becomes valid only in the limit of infinitesimal spacetime gradients), causality violations already fall outside the range of validity of the theory (even in the Newtonian limit!). In addition, \citet{Kost2000} remarked that the diffusion-type equations have a preferred reference frame by construction, namely the reference frame of the material. Thus, the fact that the diffusion equation can be solved only in this particular frame is by no means a good reason for considering it wrong.

We adopt the interpretation of \citet{Kost2000}. In fact, we may think of the diffusion equation as arising from a symmetry-breaking \cite{Ojima1986}, where the symmetry group is the proper orthochronous Lorentz group $SO^+(3,1)$: certainly the fundamental equations governing the system must be fully covariant under the action of $SO^+(3,1)$, but we are expanding them around a particular reference state (the system's equilibrium state) that is not invariant under the full action of $SO^+(3,1)$, as it has a well defined center-of-mass four-velocity.
From this point of view, there is no reason to insist on the well-posedness of the initial value problem in an arbitrary frame, as covariance has been broken at the level of the fundamental assumptions underlying a model for diffusion.

We believe that one of the the deepest insights on the issue of the diffusion equation being ``right'' or ``wrong'' has been given by \citet{Geroch95}. He proposed that we should think of relation \eqref{zentratre}, not as an equation to be solved for arbitrary initial conditions, but rather as the observation that the difference between its two sides is too small to be measurable in many physical states. This point of view is fully confirmed by kinetic theory and by the work of \citet{LindblomRelaxation1996}. In fact, as we have verified in subsection \ref{degmuiono}, the relations \eqref{hhhh} and \eqref{FFouRRieRRzxub} emerge only as approximate large-time slow-limit features of the solutions of \eqref{inhomoGG}. There is no reason to believe that they should produce a well-posed initial value problem of their own, as we already know that they become valid only on late times ($t \gg \tau$).  

In conclusion, if equation \eqref{zentratre} is valid on a given domain, then it is valid in the sense of \citet{Geroch95} in any reference frame. On the other hand,  it can be used to generate a well-posed initial value problem only in the preferred reference frame of the material and, therefore, in this frame it can be solved with finite precision methods. The corresponding solutions are, then, approximately causal and have physical significance within the range of validity of the theory ($\omega \rightarrow 0$). 
The problems usually attributed to the Navier-Stokes-Fourier approach appear again in all those situations in which a global rest frame of the material does not exist. This is the case for example of a fast-rotating relativistic body (such as a millisecond pulsar) or, more simply, a generic inhomogeneous fluid motion (in the non-linear regime). This possibility is one of the driving forces for developing alternative theories, which will be the topic of the next section. 

It is, finally, important to remark that the search for a hyperbolic alternative to Navier-Stokes-Fourier is a problem of interest also in a Newtonian context. It naturally arises whenever one wishes to model, using a hydrodynamic approach, the dynamics of systems which are not in the slow limit, e.g. the early phase ($t \lesssim \tau$) of \eqref{hoftitty}. Contrarily to what intuition may suggest, abandoning the slow limit does not necessarily imply that a hydrodynamic description becomes unfeasible \cite{Jou_Extended,AnilePavon1998,Herrera2001}, as we explicitly verified in subsection \ref{JonasCurioso} with our simple toy-model.

\section{Two approaches to solve the instability problem}

In this section we present the two most promising ways to overcome the instability problem of the Navier-Stokes-Fourier approach: the frame-stabilised first-order approach and the second-order approach. The latter will constitute the starting point of UEIT.

\subsection{A thermodynamic view on the instability problem}

In order to have a complete picture of the origin of the boost-generated spurious modes, we need to answer a final question: why does it happen that $\text{Im}(\omega_B)>0 \,$? In fact, this is both the origin of the instability discussed by \citet{Hiscock_Insatibility_first_order}, and the cause that makes the initial value problem ill-posed. Within a Lagrangian framework, a deep mathematical explanation of the instability has been given by \citet{TorrieriIS2016}, who showed that the action of Navier-Stokes does not have a global minimum. However, here we will focus on a more recent thermodynamic argument. 

\begin{figure}
\begin{center}
	\includegraphics[width=0.49\textwidth]{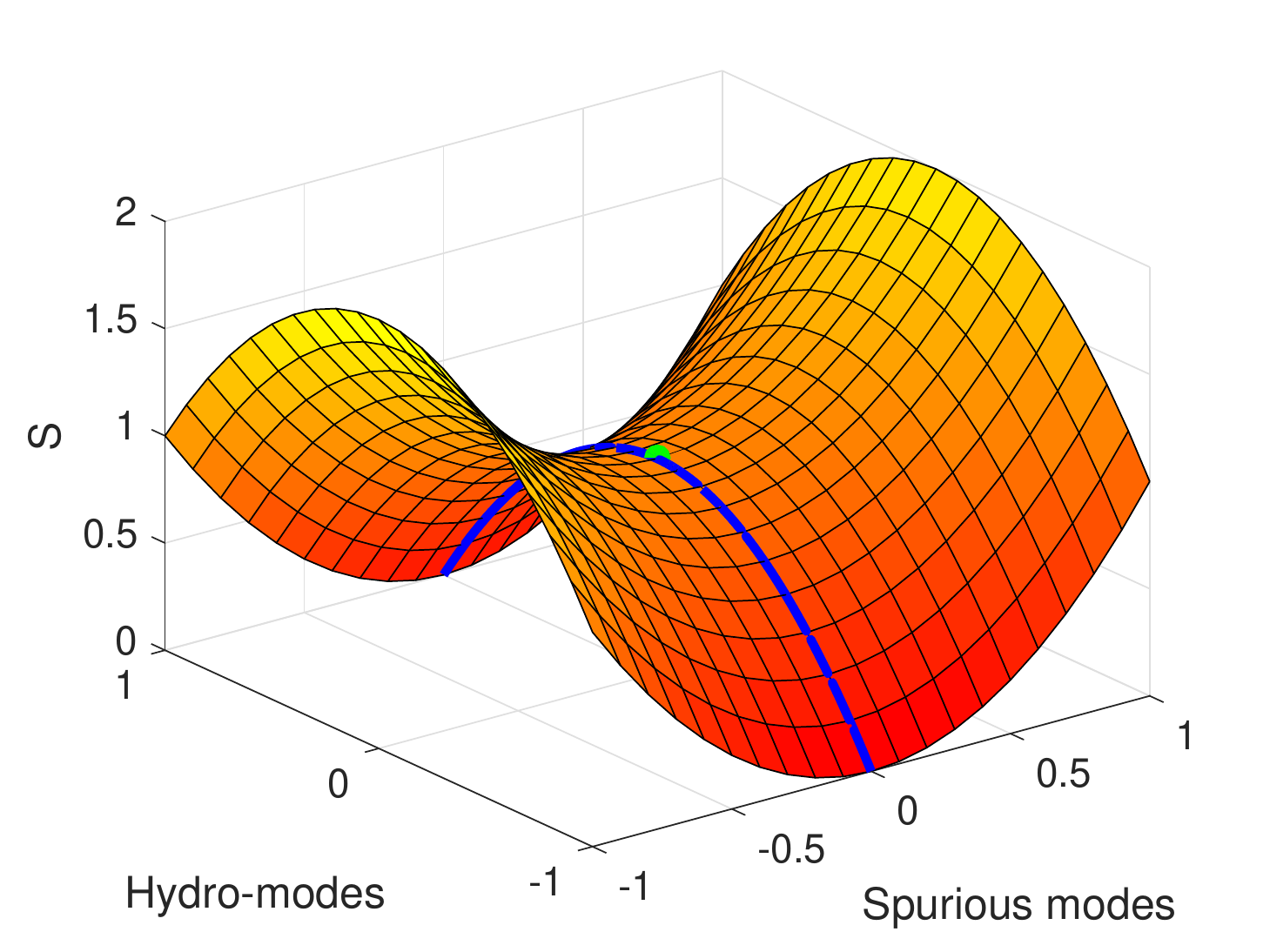}
	\caption{Sketch of the entropy $S$ along the modes of a generic first-order theory (arbitrary units). The green saddle point is the equilibrium state (i.e. the homogeneous perfect fluid state) of a generic Navier-Stokes-Fourier model, the blue line marks the states accessible by the Newtonian limit of the model. Along the Hydro-modes the entropy decreases: these modes are damped if we impose the validity of the second law. Along the spurious modes the entropy grows: the second law forces them to grow indefinitely, originating the instability.}
	\label{fig:First_order_entropy}
	\end{center}
\end{figure}

A simple answer to the question of what causes the instability is that the total entropy of the system  grows as we increase the amplitude of the spurious gapped modes \citep{GavassinoLyapunov_2020}.
This mechanism is summarised in Figure \ref{fig:First_order_entropy}. The existence of the spurious modes converts the equilibrium state from the absolute maximum of the entropy into a saddle point. As a consequence, if the second law of thermodynamics is imposed as an exact mathematical constraint (i.e. a constraint valid at all the orders), the spurious modes  have a tendency to grow. From this point of view, the positive sign of $\Im(\omega_B)$ is a consequence of the fact that also the spurious gapped modes contribute to the the increase of entropy in time. In appendix \ref{entropyplots}, we consider the example of the diffusion equation, which has not been discussed explicitly in \citet{GavassinoLyapunov_2020}.

This new insight immediately points towards two possible solutions of the problem. The first consists of realising that, since the spurious modes are unphysical, so is the entropy profile along them. Hence, there is no reason to constrain the entropy production to be non-negative along them: this is the path leading to the frame-stabilised first-order theories. The second solution consists of trying to improve the form of the entropy, by including additional pieces which restore the presence of an absolute maximum, instead of a saddle point. This leads to the second-order theories.

\subsection{The frame-stabilised first-order approach}

The frame-stabilised first-order theory, that we will call Bemfica-Disconzi-Noronha-Kovtun (BDNK) theory, is a causal and strongly hyperbolic hydrodynamic model for relativistic dissipation  \citep{BemficaDNDefinitivo2020}. 
The BDNK approach is based on the fundamental postulate (upon which also Navier-Stokes is based) that the fields of the theory are exactly the same as those of the perfect fluid \citep{Kovtun2019}:
\begin{equation}\label{hjkub}
(\varphi_i) = (u^\sigma,\Theta,\mu) \, ,
\end{equation}
see equation \eqref{kovtunN}. This implies that also this theory is assumed to have physical applicability only in the slow regime. The constitutive relations are expanded in the derivatives of the fields and truncated at the first-order as
\begin{equation}\label{functionalkovtun}
\begin{split}
& T^{\nu \rho} = T^{\nu \rho}_{\text{pf}}+ \mathfrak{T}^{ \nu \rho \sigma i}  \, \nabla_\sigma \varphi_i \\
& s^\nu = s_{\text{pf}}^\nu - \Theta^{-1} (u_\rho \mathfrak{T}^{ \nu \rho  \sigma i} + \mu \mathfrak{N}^{ \nu  \sigma i}) \,  \nabla_\sigma \varphi_i \\
& n^\nu = n^\nu_{\text{pf}}+ \mathfrak{N}^{\nu  \sigma i}  \, \nabla_\sigma \varphi_i \,. \\
\end{split}
\end{equation}
The subscript ``pf'' refers to the perfect-fluid constitutive relations (which do not involve derivatives) and we are using the Einstein summation convention for the repeated index $i=u,\Theta,\mu$. The tensor fields
\begin{equation}
\mathfrak{T}^{ \nu \rho \sigma i}  = \mathfrak{T}^{ (\nu \rho) \sigma i}  (u^\sigma,\Theta,\mu)   \quad \quad \mathfrak{N}^{\nu  \sigma i} = \mathfrak{N}^{\nu  \sigma i} (u^\sigma,\Theta,\mu)
\end{equation}
are the first-order coefficients of the derivative expansion and carry an additional hidden index in the case $i = u$. 
The constitutive relations \eqref{functionalkovtun} are the most general possible, given the choice of fields \eqref{hjkub} and the first-order truncation in the derivatives; no thermodynamic or geometrical argument is used to impose constraints on $ \mathfrak{T}^{ \nu \rho \sigma i}$ and $\mathfrak{N}^{\nu  \sigma i} $. In this sense, the BDNK approach makes manifest the effective field theory nature of hydrodynamics, according to which $u^\sigma,\Theta,\mu$ are just evocative names for the dynamical fields upon which the model is built: they do not need to carry any deep physical meaning out of equilibrium.

The hydrodynamic equations are simply the conservation laws \eqref{conservotutto}, which, given the constitutive relations above, are clearly of the second order. Hence, there are $4+1=5$ second-order equations for $3+1+1=5$ independent functions (the independent components of the fundamental fields), producing a system with
$\mathfrak{D}=2\times 5=10$ degrees of freedom. The fact that $\mathfrak{D}$ is exactly twice the number of degrees of freedom of Navier-Stokes generates many spurious gapped modes with no Navier-Stokes analogue \cite{Kovtun2019}.

For these theories the total entropy computed from $s^\nu$  still behaves as in Figure \ref{fig:First_order_entropy}. However, since along the gapped modes the time-derivatives are necessarily not infinitesimal (due to the presence of the gap that forces $\omega$ to be finite), the derivative expansion is no longer applicable. This, apart from confirming that the gapped modes are unphysical in a first-order approach, shows us that the entropy current given in \eqref{functionalkovtun} is necessarily incomplete along these modes. Hence, there is no reason for imposing the validity of the condition $\nabla_\nu s^\nu \geq 0$ also along the gapped modes (and more in general for large gradients).

It turns out that leaving $ \mathfrak{T}^{ \nu \rho \sigma i}$ and $\mathfrak{N}^{\nu  \sigma i} $ completely free (without imposing any thermodynamical or geometrical constraint on them) allows one to regulate them in a way that both the gapped and the Hydro-modes decay with time. This enforces the stability of the theory \citep{Bemfica2019_conformal1,Kovtun2019,BemficaDNDefinitivo2020} at the expenses of allowing for $\nabla_\nu s^\nu<0$ when gradients are large \citep{GavassinoLyapunov_2020,Shokri2020}. This a posteriori regulation of $ \mathfrak{T}^{ \nu \rho \sigma i}$ and $\mathfrak{N}^{\nu  \sigma i} $ is commonly referred to as choice of  \textit{hydrodynamic} frame (not to be confused with \textit{reference} frame \cite{Kovtun2019}), hence the name \textit{frame-stabilised theory}.

In this way, we end up with a relativistic dissipative first-order theory that allows us to set the initial conditions arbitrarily. In fact, given a generic initial state, we only need to integrate the equations for a characteristic time-scale (given by the inverse of the imaginary part of the gaps), so that the gapped modes have time to decay and the system relaxes to a physical solution. This mechanism is similar to the Relaxation Effect described by \citet{LindblomRelaxation1996}. 
The philosophical difference is that, while in the case of \citet{LindblomRelaxation1996} the hydrodynamic theory has the ambition of providing a consistent description also of the early-time evolution (so, in some sense, the aim is to develop a theory that is more fundamental than Navier-Stokes), in the first-order BDNK approach the early transient is considered to be unphysical.

\subsection{The second-order approach}

\begin{figure}
\begin{center}
	\includegraphics[width=0.49\textwidth]{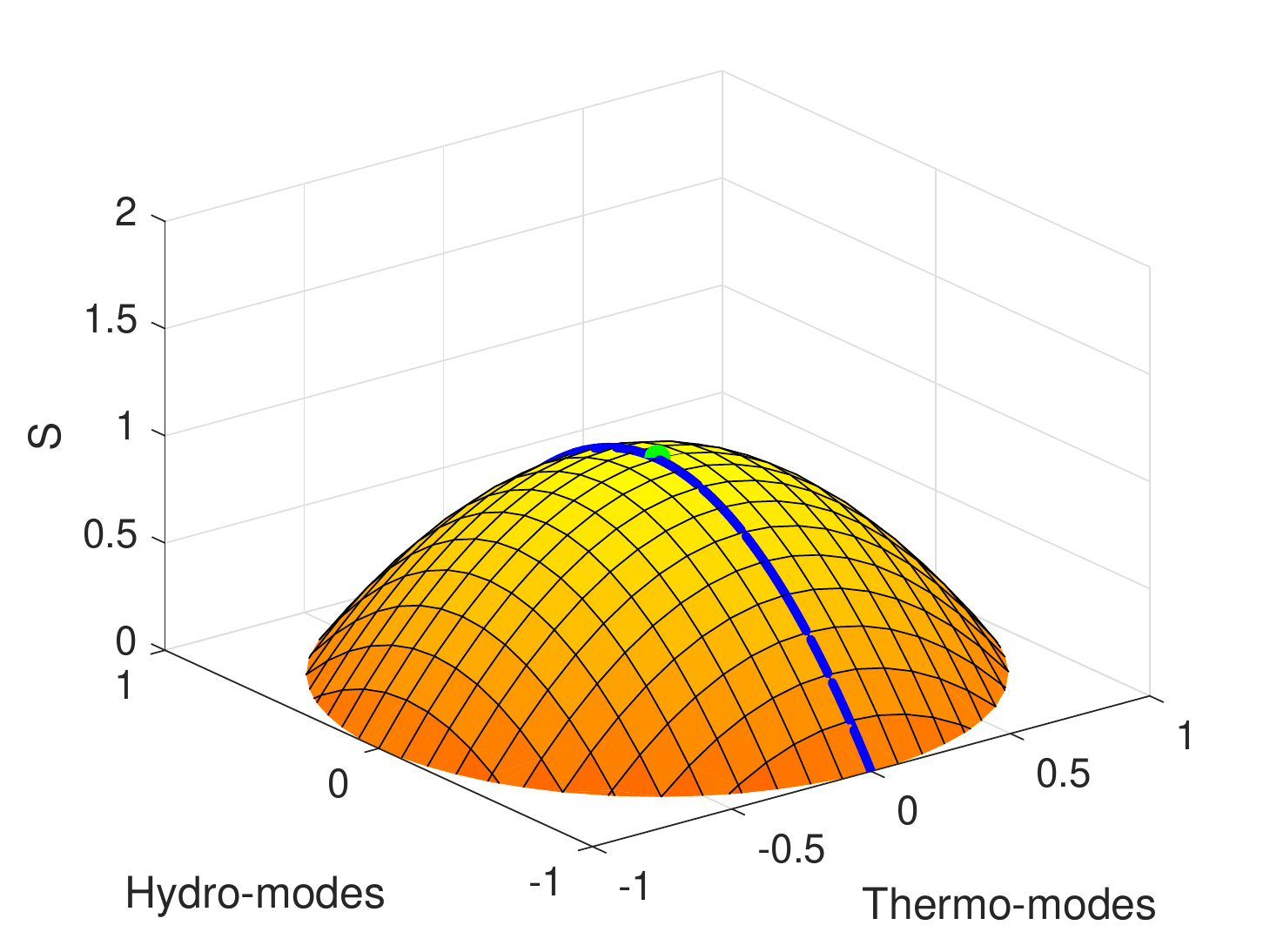}
	\caption{The physical entropy $S$ of a generic fluid (arbitrary units). 
	The absolute maximum (the green point) is the equilibrium state, namely the homogeneous perfect fluid state. The blue line crosses the states accessible by the Navier-Stokes theory. Any deviation from equilibrium reduces the entropy and therefore must decay when the second law is imposed.}
	\label{fig:Second_order_entropy}
	\end{center}
\end{figure}

Let us consider again subsection \ref{degmuiono}, and let us focus on the complete kinetic equation \eqref{inhomoGG}. As we said, if we can neglect the first term on the left-hand side we are left with \eqref{hhhh}. From this we recover the Navier-Stokes-Fourier approach, which emerges as the slow-limit, large-scale, late-time behaviour of the fluid. If, on the other hand, we neglect the right-hand side of \eqref{inhomoGG}, we are left with the solution
\begin{equation}
\dfrac{\tau}{p^0} \,  p^\nu \partial_\nu h + h =0 
\qquad \Rightarrow \qquad
h(t) =h(0)e^{-t/\tau} \, ,
\end{equation}
along the geodesic path \eqref{geoddetica}. We have recovered the early-time relaxation typical of a non-equilibrium thermodynamic system. Therefore, a ``realistic'' fluid system, prepared with arbitrary initial conditions (but still close to equilibrium), is expected to exhibit two distinct behaviours: an initial fast local thermodynamic evolution of the fluid elements, whose dynamics can only be captured by some Thermo-modes, and a late-time hydrodynamic evolution of the Hydro-modes. 
With this in mind, it seems natural to seek a complete description of the fluid that unites, within a single formalism, non-equilibrium thermodynamics and dissipative hydrodynamics. These two theories should emerge as limiting behaviours of such a unified formalism when only one of the two classes of modes is considered:
\begin{align*}
 \text{Thermo-modes (gapped)} \, & \rightarrow &\text{Non-equilibrium}   \quad  & \\ 
                                 & \,          & \, \text{thermodynamics}  \quad  &
  \\
   & & &
  \\
 \text{Hydro-modes (gapless)} \, & \rightarrow & \text{Navier-Stokes-Fourier} & \\ 
                                 & \,          & \, \text{approach}  \quad \quad \quad &
\end{align*}
Clearly, for a consistent connection with non-equilibrium thermodynamics to be possible, we have to expand the entropy at least to the second order in any deviation from equilibrium, by replacing the non-realistic profile of Figure \ref{fig:First_order_entropy} (which is the direct product of a first-order truncation of the entropy current) with the profile in  Figure \ref{fig:Second_order_entropy} (which is what we must obtain, for small deviations from equilibrium, if we truncate at the second order).

The aforementioned reasoning constitutes the rationale of Extended Irreversible Thermodynamics \citep{Jou_Extended}, which naturally leads to the formulation of the second-order theory of \citet{Israel_Stewart_1979}. The structure of this hydrodynamic model, as a classical effective field theory, will be discussed later. For the time being, we only mention that the theory treats the dissipative fluxes, such as heat flux and viscous stresses, as the only non-equilibrium thermodynamic variables, which are subject to their own relaxation processes, producing the Thermo-modes. 
The downside of this approach is that it necessarily leads to a hydrodynamic theory with $\mathfrak{D}=14$ degrees of freedom. Because of this, the second-order theory of \citet{Israel_Stewart_1979} can not have the same universality property of non-equilibrium thermodynamics, which can, in principle, deal also with a larger number of independent thermodynamic variables. As a consequence, standard Extended Irreversible Thermodynamics (EIT) cannot be used to capture the dynamics of many exotic fluids, such as those that arise from  strongly coupled gauge theories \citep{Denicol_Relaxation_2011,Heller2014}.

However, the idea of EIT of considering heat conduction and viscosity as arising from the dynamical coupling between the internal evolution of non-equilibrium thermodynamic variables and the hydrodynamic evolution is the starting point of a formulation of a more universal non-equilibrium thermodynamic theory for relativistic fluids, which contains EIT as a particular case \citep{noto_rel,BulkGavassino}. The second part of this review will be devoted to presenting this more general theory, which we will refer to as Unified Extended Irreversible Thermodynamics, that, in extreme synthesis, is just non-equilibrium thermodynamics applied to the local fluid elements.


\section{Unified Extended Irreversible Thermodynamics}

We introduce the principles of Unified Exteded Irreversible Thermodynamics (UEIT) and list its fundamental features. Many of the following ideas already appear (in different forms) in the literature, but here we aim at giving a more systematic presentation, referring to the original works whenever possible.

By UEIT we mean a description of dissipative fluids which extends the language of non-equilibrium thermodynamics to a hydrodynamic context. From a thermodynamic perspective, a fluid is just a large thermodynamic system which may or may not be in a state of global thermodynamic equilibrium. Dissipation is the macroscopic manifestation of all the processes driving the system towards the maximum entropy state. 

The point of view of UEIT (which is supported by kinetic theory, see subsection \ref{degmuiono}) is that viscosity and heat conduction arise because fluid elements can be driven out of equilibrium during the motion generated by the presence of gradients: dissipation occurs as the internal processes act to bring each element back to equilibrium. As a consequence, the relationship between the dissipative fluxes and the gradients, which in the Navier-Stokes-Fourier approach is introduced at the level of the constitutive relations, is dynamical in UEIT, resulting from a coupling between the evolution of the internal (non-equilibrium) thermodynamic degrees of freedom of a fluid element and the gradients.
  
\subsection{Basics of non-equilibrium thermodynamics}
\label{basilKs}

In equilibrium thermodynamics, the state at a given time is defined by assigning some state-variables which represent the constants of motion of the system\footnote{
We do not include parameters like the volume in the state variables because they are external conditions that we assume fixed.}, 
like the total energy $U$ and any conserved charge $N$ (for simplicity we set the total momentum to zero). In non-equilibrium thermodynamics \citep{prigoginebook}, the number of state-variables is enlarged to include some additional parameters $\xi_1,\xi_2,...$ , so that the state-space is now equipped with the coordinate system
\begin{equation}\label{hikpfijmv}
(X_i) = (U,N,\xi_1,\xi_2,...) \, .
\end{equation}  
The variables $\xi_1,\xi_2,...$ are some macroscopic quantities which are not conserved and that can have a  value different from the one imposed by assuming full thermodynamic equilibrium.

Each point of the state-space is interpreted as a macrostate, namely as the  ensemble of all the quantum states of the system in which the quantum observables $(\hat{X}_i)$ take the average values $(X_i)$ within a small assigned uncertainty \citep{peliti_book}. Given the number of these microscopic states $\Gamma(X_i)$, the entropy of the macrostate is  
\begin{equation}\label{jkbnm}
S(X_i) = \ln \Gamma(X_i) \, .
\end{equation}  
Equation \eqref{jkbnm} generalizes the concept of equation of state when we are out of equilibrium. 

The description of an isolated non-equilibrium thermodynamic system is, then, completed by assuming some dynamical equations of motion (in the thermodynamic limit random fluctuations are neglected) for the point in the state-space of the form
\begin{equation}\label{rfrog}
\dot{X}_h = \mathcal{F}_h (X_i) \, ,
\end{equation}
which must be consistent with the conservation laws
\begin{equation}\label{qwuiopv}
\dot{U}=0   \quad \quad   \dot{N}=0
\end{equation}
and with the second law of thermodynamics
\begin{equation}\label{lasecleggeggerea}
\dot{S} \geq 0 \, .
\end{equation}
This scheme, defined by \eqref{hikpfijmv}-\eqref{lasecleggeggerea}, is common in most formulations of non-equilibrium thermodynamics \citep{prigoginebook}. Then, for given values of $U$ and $N$, it is possible to identify the \textit{thermodynamic equilibrium state}
$(U,N,\xi_1^{\text{eq}}(U,N),\xi_2^{\text{eq}}(U,N),...)$
by imposing the maximal-entropy condition
\begin{equation}
\dfrac{\partial S}{\partial \xi_i} \bigg|_{U,N} =0   \quad \quad   \dfrac{\partial^2 S}{\partial \xi_i \partial \xi_h} \bigg|_{U,N} < 0 \, . 
\end{equation}
This thermodynamic equilibrium state is clearly also an equilibrium state of the dynamics defined by \eqref{rfrog} and, since $S$ is a Lyapunov function of the system, it is also necessarily Lyapunov-stable \citep{lasalle1961stability}. 

\subsection{Incomplete equilibrium VS relaxation-time approximation}\label{laparlata}

It is important to understand under which conditions the scheme defined by \eqref{hikpfijmv}-\eqref{lasecleggeggerea} is applicable. 
While the constructions in \eqref{hikpfijmv} and \eqref{jkbnm} are always possible at the formal level, a critical assumption hides in \eqref{rfrog}: the dynamics defined by \eqref{rfrog} should hold for the overwhelming majority of the microstates that realize the macrostate $(X_i)$, so that we can attribute a unique macroscopic evolution to the ensemble itself. This amounts to postulating that if we pick up randomly a quantum state belonging to the class identified by $(X_i)$ and we trace its evolution (governed by the Hamilton operator $\hat{H}$), then \eqref{rfrog} is  respected almost-surely (within a certain level of approximation). There are two possible situations in which this condition is achieved. 

The most common one is when the quantities $\xi_i$ (although being not conserved) are \textit{quasi-constants of motion}, meaning that they evolve on a time-scale $\tau_M$ which is much larger than the microscopic time-scale $\tau_m$ on which the particles' individual momenta are randomised. 
If this is the case,  the variables $\xi_i$ are almost constant  on a time-scale $\tau_m$  and the state \eqref{hikpfijmv} can be seen as an \textit{incomplete equilibrium state} \citep{landau5}, 
or state of \textit{quasi-equilibrium}. 
Then, assuming that the variables $\xi_i$ are the only quasi-constants of motion, a time-average over $\tau_m$ coincides with a statistical average over the ensemble represented by the macrostate $(X_i)$. Thus, equation \eqref{rfrog}, interpreted as a time-averaged equation, holds for almost any microstate in the ensemble.
In chemical kinetics, for example, the variables $\xi_i$ are non-conserved particle numbers or reaction coordinates, which are assumed to evolve more slowly than the typical timescale on which the particles' momenta relax to the Maxwell-Boltzmann distribution \citep{prigoginebook}. Another example of time-scale separation are two-temperature systems, which will be discussed in section \ref{TwoTemper}.

An alternative setting in which \eqref{rfrog} is respected is when  the evolution equations for the observables $X_i$, neglecting small fluctuations, decouple from the evolution of the remaining (practically infinite) degrees of freedom of the system. In this case, for the overwhelming majority of the microstates in the ensemble, equations \eqref{rfrog} are enforced by the microscopic dynamics itself, even without imposing any time-average. 
For example, this is the case when the relaxation-time approximation of kinetic theory
holds.
The crucial aspect of this approximation is that each observable follows a decay rule which is independent from the exact value of most of the other observables. This allows us to decouple some relevant dynamical variables from the rest of the degrees of freedom, producing the effective dynamics in \eqref{rfrog}.

This second possibility constitutes the rationale of the Grad 14-moment approximation \cite{Israel_Stewart_1979}, as well as of any closure scheme \cite{AnileRadiazion1992}, and is the standard setting considered in the framework of the original \textit{Extended Irreversible Thermodynamics}, where it is assumed that all the complexity of the Boltzmann collision integral is captured by a simpler relaxation process involving the dissipative fluxes as the \textit{only} relevant variables \citep{cattaneo1958,Jou_Extended}.


\subsection{The fundamental principles of UEIT}
\label{fondamentaloniRADICEdiA}

The intuitive idea of UEIT is to interpret hydrodynamics as a non-equilibrium thermodynamic theory that also allows to explicitly track  the inhomogeneities of the fluid. Following this philosophy, it is natural to promote the variables $X_i$ (which describe the global properties of a system) to fields $\varphi_i$ (carrying statistical information about the local state of matter at every point): to move from non-equilibrium thermodynamics to hydrodynamics, we just have to make the formal identification
\begin{equation}\label{analogyY}
(X_i) \longrightarrow (\varphi_i) \, .
\end{equation}
%
%
There is, however, an important subtlety. The variables $(X_i)$ can be used to uniquely specify all the relevant properties of the system at a given time $t$. In fact, each choice of the values  $(X_i)$  specifies a unique ensemble at a given time $t$, which may then be used to compute the expectation value $A(X_i)$ of any quantum observable $\hat{A}(t)$. This implies that, if we want to interpret $\varphi_i$ as a ``local analogue'' of $X_i$, the fields $\varphi_i$ should specify all the relevant thermodynamic properties of the system at a given \textit{space-time} point $x\in \mathcal{M}$. 
This means that, if we know the value of all the fields $\varphi_i$ at $x$, we also know the local macrostate of the fluid in a small neighbourhood of $x$ and we can, therefore, compute the average of any local observable.
We are, then, led to restrict our attention to the constitutive relations of the kind 
\begin{equation}\label{constitutiverelationszzzUEIT}
 T^{\nu \rho} = T^{\nu \rho}(\varphi_i) \quad \quad\quad
 s^\nu = s^\nu (\varphi_i) \quad \quad \quad
 n^\nu = n^\nu (\varphi_i) \, .
\end{equation}
This implies that a fluid element in the neighbourhood of $x$ should be considered as a small non-equilibrium thermodynamic system, where $\varphi_i(x)$ represents a set of values $X_i$ for this element: the fields $\varphi_i$ play the role of local (non-equilibrium) thermodynamic variables of the fluid elements. 
The bridge with statistical mechanics is, thus, ensured by construction.

This transition from global to local by promoting the thermodynamic variables to fields is nothing but the generalization to the non-equilibrium case of what is usually done to move from equilibrium thermodynamics to perfect fluids \citep{MTW_book}. This description is expected to break down in the limit of large accelerations due to quantum effects \cite{BecattiniQuantumCorrections2015}, but as long as a perfect-fluid description is feasible in equilibrium, it should be possible to model non-equilibrium corrections with UEIT.

Regarding the total entropy of the fluid, in UEIT $S$ still has the same interpretation as in \eqref{jkbnm}; moreover, it can be computed on a spacelike hypersurface $\Sigma$ as
\begin{equation}\label{INT1}
S[\Sigma,\varphi_i] = -\int_\Sigma s^\nu(\varphi_i) \, d\Sigma_\nu
\end{equation}
and it is completely determined solely by the value of the fields on $\Sigma$. 

Let us, now, focus on the hydrodynamic equations. The dynamic equations \eqref{rfrog} are of the first order in time, due to the fact that the variables $(X_i)$ define the initial conditions completely (see subsection \ref{basilKs}): the $(X_i)$ must be the degrees of freedom of the thermodynamic model. Therefore,  we impose that the hydrodynamic equations in UEIT are of the first-order in time. 
In this way, the fields $(\varphi_i)$ are automatically the degrees of freedom of the theory. Furthermore, by imposing that this remains true in every reference frame, we are forced to rule out higher order derivatives in space, so that the hydrodynamic equations must take the form
\begin{equation}\label{hydrouzZZUEIT}
\mathfrak{F}_h(\varphi_i,\nabla_\sigma \varphi_i)=0 \, . 
\end{equation} 
In the case of a simple fluid, one can also require that all these equations must be dynamical (i.e. involve derivatives in time in every reference frame) and that they cannot be converted into constraints on the initial conditions, implying  that the number of algebraically independent components of the fields coincides with the number $\mathfrak{D}$ of hydrodynamic degrees of freedom (this condition cannot be imposed in a superfluid context, due to the irrationality constraint of the superfluid momentum \cite{carter92,cool1995}).

As a consistency check, we  see that, when we require the validity of \eqref{conservotutto}, by using the constitutive relations \eqref{constitutiverelationszzzUEIT} we obtain equations of the form \eqref{hydrouzZZUEIT}. This shows the formal compatibility of these hydrodynamic equations with the conservation laws.

The final piece of the puzzle we need to analyse is the validity of the second law of thermodynamics. As we anticipated, the dynamical equations \eqref{rfrog} neglect fluctuations. We can, thus, use them as long as the number of particles in each volume element is sufficiently large that we can neglect the uncertainties\footnote{
This depends on the resolution we want to achieve. If this was not the case, we would need to replace the ordinary dynamical equations with stochastic equations, see e.g. \citet{TorrieriCrooks2020}.
}.
This is the local version of the usual thermodynamic limit: within this assumption, the equations \eqref{hydrouzZZUEIT} must ensure the validity of the second law as an exact condition \citep{huang_book}. 
Formally, we should require that
\begin{equation}\label{Positto}
\nabla_\nu \big[ s^\nu(\varphi_i) \big] := \sigma (\varphi_i , \nabla_\sigma \varphi_i) \geq 0
\end{equation}
for any possible choice of $(\varphi_i , \nabla_\sigma \varphi_i)$ satisfying the conditions \eqref{hydrouzZZUEIT} in the point under consideration. Only in this way we can guarantee that also \eqref{lasecleggeggerea} is automatically respected \cite{Hishcock1983} for an arbitrary initial configuration. 

Collecting together all these ideas, the principles of UEIT are summarized into three statements:  
\begin{enumerate}
\item the constitutive relations do not involve derivatives of the fields,
\item the hydrodynamic equations are all of the first order both in space and in time,
\item the second law is enforced for any initial condition.
\end{enumerate}
Any hydrodynamic model that is presented in a form that respects the foregoing requirements is formally a UEIT-theory and should admit, in principle, a direct bridge with non-equilibrium thermodynamics.

\subsection{The near-equilibrium and the slow-evolution assumptions in UEIT}

In the formulation of the basic principles of UEIT no explicit reference has been made to the near-equilibrium assumption. 
In fact, the condition of quasi-equilibrium introduced in subsection \ref{basilKs} does not necessarily imply a near-equilibrium condition. In fact, the expression \textit{quasi-equilibrium} refers to the fact that, due to a time-scale separation, we can describe the system referring to the ensemble $(X_i)$. On the other hand, the \textit{near-equilibrium} assumption is the requirement that the average value of any observable is close to the equilibrium value. A system in quasi-equilibrium is also near-equilibrium if all the differences $\xi_i-\xi_i^{\text{eq}}$ are small. 

It is important to remark that in non-equilibrium thermodynamics the near-equilibrium condition is, typically, not necessary. For example, in chemistry one may 
%
model a far from equilibrium reaction  by using only a finite number of reaction coordinates as free variables.
Therefore, UEIT is, at least in principle, non-perturbative and frame-independent (we refer here to the concept of \textit{hydrodynamic frame} described by \citet{Kovtun2019}).

Clearly, the near-equilibrium hypothesis can still be invoked for some specific UEIT models, or it may even be a practical necessity, as there are situations where the quasi-equilibrium assumption (or the relaxation-time approximation) holds only in a neighbourhood of equilibrium.  

Regarding the slow limit, its role within  the UEIT framework is subtle. Since the original purpose of EIT is to avoid the formal introduction of this limit, the slow-evolution assumption is not a fundamental assumption of UEIT either. However, as mentioned in subsection \ref{laparlata}, in many contexts the non-equilibrium thermodynamic description is valid only on time-scales which are much larger than the typical collision time $\tau_m$. Therefore, the slow-evolution assumption in UEIT may be necessary for the predictions of a particular hydrodynamic model to be reliable, even though it is not invoked as a fundamental postulate of the theory. In any case, the important point is that, as we are going to show, the structure of UEIT guarantees that the slow limit does not compromise the stability of the equilibrium and can not produce spurious gapped modes.


\subsection{The equilibrium in UEIT must be Lyapunov stable}\label{LyiYup}

To study the stability of the UEIT equilibrium state, let us assume a fixed background spacetime that admits a global time-like and future-oriented Killing field $K^\nu$. Then,  we also have a conserved current $J^\nu$,
\begin{equation}
J^\nu := -T^{\nu \rho}K_\rho
\qquad \qquad \quad
\nabla_\nu J^\nu =0 \, .
\end{equation}  
Assuming that the fluid has a finite spatial extension, the Gauss theorem ensures that the quantity \cite{carroll_2019}
\begin{equation}\label{INT2}
U[\Sigma,\varphi_i] = \int_\Sigma T^{\nu \rho}(\varphi_i) \, K_\rho \, d\Sigma_\nu
\end{equation}
does not depend on the choice of the spacelike hypersurface $\Sigma$, provided that this extends to infinity \citep{MTW_book}. The same holds for the quantity
\begin{equation}\label{INT3}
N[\Sigma,\varphi_i] = - \int_\Sigma n^\nu(\varphi_i) \, d\Sigma_\nu \, .
\end{equation}
We may identify the quantities $U$ and $N$ with the total energy\footnote{
Despite the formal similarity, the total energy $U$ is not the Komar mass of the system. In fact, the conservation of the  Komar mass is a feature of systems immersed in a dynamic spacetime, while we are working with a fixed background spacetime.} 
associated with the Killing vector $K$ and the particle (minus anti-particle) number of the fluid. For simplicity, we assume that there are no other Killing vector fields. 

Now let us also assume that we can introduce a global coordinate system $(t,x^1,x^2,x^3)$ such that $K = \partial_t$.
The surfaces $\Sigma(t)$ at constant time define a global foliation of the spacetime and we can define $S(t)$, $U(t)$ and $N(t)$ as the integrals in \eqref{INT1}, \eqref{INT2} and \eqref{INT3} for $\Sigma=\Sigma(t)$. Recalling \eqref{Positto}, we have that
\begin{equation}
\label{ilGhoulGourmet}
\dfrac{dS}{dt} \geq 0   \qquad \qquad   
\dfrac{dU}{dt}= 0   \qquad \qquad    
\dfrac{dN}{dt} =0 \, .
\end{equation}
Therefore, if we interpret the whole fluid as an inhomogeneous thermodynamic system, we see that it obeys the same global laws \eqref{qwuiopv} and \eqref{lasecleggeggerea} as in standard non-equilibrium thermodynamics.

Now, let us move to  the hydrodynamic equations in \eqref{hydrouzZZUEIT}. By working in the coordinate system $(t,x^j)$, we can recast \eqref{hydrouzZZUEIT} in a form that is the analogue of \eqref{rfrog} under the replacement \eqref{analogyY}, namely\footnote{
There is no explicit dependence of $\mathcal{F}_h$ on $t$ because the Killing condition for the vector field $K$ in this coordinate system reduces to $\partial_t g_{\nu \rho}=0 $, so that $g_{\nu \rho}=g_{\nu \rho}(x^j)$, see e.g.  \cite{carroll_2019}.
}
\begin{equation}\label{systemMKL}
\partial_t \, \varphi_h \, = \, 
\mathcal{F}_h ( \, \varphi_i \,  , \,  \partial_j \varphi_i  \,  ,  \,  x^j \, ) \, .
\end{equation}
This system in \eqref{systemMKL} describes an autonomous (i.e. time-translation invariant) dynamical system, whose state-space is a space of functions from $\mathbb{R}^3$ to $\mathbb{R}^\mathfrak{D}$, where $\mathfrak{D}$ is the number of degrees of freedom of the model, see subsection \eqref{vrif}. 
Casting the hydrodynamic equations  as in \eqref{systemMKL} is, at least in principle, always possible because all the equations \eqref{hydrouzZZUEIT} were assumed to be independent and dynamical (so we can always identify a term with a time derivative).

We can also rewrite $S(t)$, $U(t)$ and $N(t)$ in the aforementioned coordinate system as \cite{carroll_2019}
\begin{equation}
\begin{split}
& S(t) = \int s^t \big(\varphi_i(t,x^j) \big) \sqrt{-g(x^j)} \, d^3 x \\
& U(t) = -\int T\indices{^t _t} \big(\varphi_i(t,x^j) \big) \sqrt{-g(x^j)} \, d^3 x \\
& N(t) = \int n^t \big(\varphi_i(t,x^j) \big)\sqrt{-g(x^j)} \, d^3 x \, . 
\end{split}
\end{equation}
%
Since the above integrals depend on time only through the fields $\varphi_i$, the global quantities  $S(t)$, $U(t)$ and $N(t)$  can be seen as functions on the state-space. Thus, in view of \eqref{ilGhoulGourmet}, the quantities $U$ and $N$ are integrals of motions of the dynamical system \eqref{systemMKL}, while $S$ is a strictly non-decreasing function of $t$. On the other hand, if the total entropy is computed consistently with microphysics, then there should be a state $(\varphi_i^{\text{eq}})$ on $\Sigma(t)$ such that
\begin{equation}
S[\Sigma(t),\varphi_i^{\text{eq}}] \, > \,  S[\Sigma(t),\varphi_i]    
\end{equation}
for any $\varphi_i \neq \varphi_i^{\text{eq}}$ such that
\begin{equation}
\begin{split}
& U[\Sigma(t),\varphi_i^{\text{eq}}] = U[\Sigma(t),\varphi_i]
\\ 
& N[\Sigma(t),\varphi_i^{\text{eq}}] = N[\Sigma(t),\varphi_i] \, .
\end{split}
\end{equation}
All these results together tell us  that $S$ is a Lyapunov function for the system. 
As a consequence, $(\varphi_i^{\text{eq}})$, which is the state of global thermodynamic equilibrium (i.e. the maximum entropy state), is also an equilibrium state of the dynamics \eqref{systemMKL} and is Lyapunov-stable.

It is interesting to note that the Lyapunov-stability follows only from the three principles of UEIT discussed in subsection \ref{fondamentaloniRADICEdiA} and from the consistency of the constitutive relations \eqref{constitutiverelationszzzUEIT} with microphysics. This is important because the details of the hydrodynamic equations, which are typically a difficult part of the model to determine, do not play any role in the final stability criterion, implying that we are  free to impose any kind of slow-evolution approximation we wish, confident that this is not going to compromise the stability of the theory.

\subsection{The importance of the Thermo-modes}

To appreciate the role of the Thermo-modes in UEIT, we have to discuss the behaviour of a generic UEIT model in the homogeneous limit. For simplicity, we now assume a Minkowski background spacetime and we restrict our attention to perfectly homogeneous states. In this case, the replacement
$(\varphi_i) \longrightarrow (X_i)$ in \eqref{analogyY}
exactly maps a UEIT model into a conventional non-equilibrium thermodynamic model. 
This creates a formal bridge with statistical mechanics, that could be used to derive the constitutive relations \eqref{constitutiverelationszzzUEIT} from microphysics.
For homogeneous states, the hydrodynamic equations assume the same form of \eqref{rfrog}, namely
\begin{equation}\label{xsabfung}
\partial_t \varphi_h = \mathcal{F}_h (\varphi_i) \, ,
\end{equation}
which is a set of $\mathfrak{D}$ ordinary differential equations. 
%
%
Since the conservation laws \eqref{conservotutto} in this homogeneous limit become
\begin{equation}
\partial_t T^{t\nu} =0   \quad \quad   \partial_t n^t =0 \, ,
\end{equation} 
the dynamics in \eqref{xsabfung} is subject to 5 independent constraints, implying that the Thermo-modes can exist if and only if
\begin{equation}\label{DEGOF}
\mathfrak{D} >5 \, .
\end{equation}
This inequality is respected by any dissipative UEIT model of a simple fluid (with one conserved charge), making the presence of Thermo-modes unavoidable.  In fact, $\mathfrak{D}=5$ is already the case for the perfect fluid, while a dissipative system must have a larger state-space to account for non-equilibrium states of the fluid elements.  

The existence of any degree of freedom above 5 corresponds, in the homogeneous limit, to the introduction of a variable of the type $\xi_i$, the  non-conserved quantities of  subsection \ref{basilKs}. In fact, the Thermo-modes arise whenever the system is able to occupy non-equilibrium homogeneous states (as discussed in subsection \ref{HydroThermo}), and the variables $\xi_i$ are coordinates on the resulting extended state-space.

In conclusion, in UEIT the number of degrees of freedom ($\mathfrak{D}>5$) is the same in every reference frame and is given by the number of independent components of the fields, so that the existence of boost-generated spurious gapped modes is automatically ruled out: every gapped mode has a clear thermodynamic interpretation as a Thermo-mode.

\subsection{Lindblom's Relaxation Effect}\label{Lindddblumz}


We briefly present the seminal work of \citet{LindblomRelaxation1996}, as it contains many ideas that are fundamental for UEIT. In doing this we will omit important technical details, as our aim is to present the physical meaning of Lindblom's results and its implications for UEIT. 

We have seen that, in order to have a dissipative theory, we must have $\mathfrak{D} >5$. In principle, we may  isolate 5 preferred degrees of freedom, whose value at each $x\in \mathcal{M}$ identifies, locally, a fiducial perfect-fluid state. 
Choosing the Eckart hydrodynamic frame \citep{Kovtun2019}, Lindblom defines the \textit{dynamical fluid fields}
\begin{equation}
(\, \phi^a \, ) := (\,u^\sigma, \,n , \, \mathcal{U} \, )
\end{equation}
as
\begin{equation}\label{proscrivuznz}
u^\sigma = n^\sigma /\sqrt{-n^\lambda n_\lambda}  
\qquad   n=-n^\sigma u_\sigma  
\qquad   \mathcal{U}=T^{\sigma \lambda} u_\sigma u_\lambda 
\end{equation}
and makes a change of variables 
\begin{equation}\label{LindblomReprentazionnze}
(\varphi_i) \longrightarrow (\phi^a,\mathbb{A}^A) \, ,
\end{equation}
where the remaining tensor fields $\mathbb{A}^A$ can be constructed freely, with the only constraint that $ \mathbb{A}^A=0 $ at local thermodynamic equilibrium (i.e. when we recover the perfect fluid on $x$). 
This condition can always be imposed, as it is always possible to perform the change of variables
\begin{equation}
\mathbb{A}^A \, \longrightarrow \, \Check{\mathbb{A}}^A := \mathbb{A}^A - \mathbb{A}^A_{\text{eq}}(\phi^a) \, .
\end{equation}
The fields $\mathbb{A}^A$ are called \textit{dissipation fields}; since $\mathfrak{D}>5$ we always have at least one dissipation field.

Now, the constitutive relation for the stress-energy tensor can  be decomposed as
\begin{equation}
T^{\nu \rho}= T^{\nu \rho}_{\text{pf}}(\phi^a) + \Delta T^{\nu \rho}(\phi^a,\mathbb{A}^A) \, ,
\end{equation}
where $T_{\text{pf}}^{\nu \rho}$ is the perfect-fluid stress-energy tensor and
\begin{equation}
\Delta T^{\nu \rho}(\phi^a,0)=0 \, ,
\end{equation}
as we know by definition that, in local thermodynamic equilibrium, the fluid must be in the perfect-fluid state.
The conservation laws \eqref{conservotutto} can play the role of 5 hydrodynamic equations (using Einstein's summation convention for the indices $a$ and $A$):
\begin{equation}
  \dfrac{\partial T^{\nu \rho}}{\partial \phi^a} \nabla_\nu \phi^a + \dfrac{\partial T^{\nu \rho}}{\partial \mathbb{A}^A} \nabla_\nu \mathbb{A}^A =0 \, ,
  \quad  
  u^\nu \nabla_\nu n+ n\nabla_\nu u^\nu  =0 \, .  
\end{equation}
The remaining equations are equal in number to the number of algebraically independent components of the dissipation fields. Hence, Lindblom makes the assumption that they can be written in the form
\begin{equation}\label{telgucz}
M\indices{^\rho _A _B} \nabla_\rho \mathbb{A}^B + \Xi_{AB} \mathbb{A}^B = - M\indices{^\rho _a _A} \nabla_\rho \phi^a \, ,
\end{equation}
where the tensor fields $M\indices{^\rho _A _B}$, $M\indices{^\rho _a _A}$ and $\Xi_{AB}$ are functions of $(\phi^a,\mathbb{A}^A)$ and should obey some conditions that are listed in \cite{LindblomRelaxation1996}. The structure of \eqref{telgucz} recalls the kinetic equation \eqref{inhomoGG}. In essence, it tells us that the dissipation fields, which in the homogeneous limit would undergo a thermodynamic relaxation towards equilibrium (left-hand side), are dynamically coupled to the gradients of the dynamical fluid fields (right-hand side). 
This can drive them out of equilibrium. This same mechanism had previously inspired also the Maxwell-Cattaneo equation, namely the second equation of \eqref{systemmum}, see \cite{cattaneo1958}. Not every UEIT theory must be governed by such a simple dynamics, but it is a common feature of many of them \citep{Geroch_Lindblom_1991_causal}.

The key result of Lindblom is that, given an initial condition with small spatial gradients, the system will eventually relax to a state in which\footnote{
Note the analogy between equations \eqref{telgucz}-\eqref{inhomoGG} and 
\eqref{abababa}-\eqref{hhhh}: in the end the dissipation fields $\mathbb{A}^A$ are expressed in terms of the gradients of the dynamical fields $\phi^a$. 
}
\begin{equation}\label{abababa}
\mathbb{A}^B \approx - (\Xi^{-1})^{BA} M\indices{^\rho _a _A} \nabla_\rho \phi^a \, .
\end{equation}
As a consequence, the deviation of the stress-energy tensor from  $T_{\text{pf}}^{\nu \rho}$ relaxes to
\begin{equation}
\Delta T^{\nu \rho} \approx \dfrac{\partial T^{\nu \rho}}{\partial \mathbb{A}^B} \mathbb{A}^B 
\approx 
- \, \dfrac{\partial T^{\nu \rho}}{\partial \mathbb{A}^B} (\Xi^{-1})^{BA} M\indices{^\rho _a _A} \nabla_\rho \phi^a \, .
\end{equation}
Hence, the Navier-Stokes behaviour emerges only after an initial relaxation, during which the dynamical equations \eqref{telgucz} are converted into the constraints \eqref{abababa}, effectively introducing gradients in the constitutive relations. Again, this clarifies why the Navier-Stokes equations do not admit a well-posed initial value problem (at least in relativity): the Navier-Stokes formulation can not admit an arbitrary initial configuration of the fields, but rather it is valid  only for those initial states that are the end states of a relaxation, after all the gapped modes of the complete UEIT theory have decayed.  

\section{The two cornerstones of UEIT}
\label{JonasLatinista}

We present two remarkable dissipative theories that are consistent with the principles of UEIT: the second-order hydrodynamic model of \citet{Israel_Stewart_1979} and the phenomenological multifluid approach of \citet{Carter_starting_point}.

\subsection{Israel-Stewart as a UEIT theory}

The hydrodynamic model of \citet{Israel_Stewart_1979} is probably the most natural example of a UEIT theory.
The fundamental assumption on which the model is built  is that we can take the conserved fluxes as fundamental fields of the theory\footnote{Since this is the fundamental premise for most of the relativistic EIT theories, we will use  the collective name  \textit{Israel-Stewart theory} to indicate  all those UEIT models that build on this assumption.}, namely
\begin{equation}\label{nTitty}
(\varphi_i) = (n^\sigma, T^{\sigma \lambda}) \, .
\end{equation}
Hence, given the symmetry of $T^{\sigma \lambda}$, the number of degrees of freedom of the  model is 
\begin{equation}
\mathfrak{D}=14 = 5+9 \, .
\end{equation}
Given the fields in \eqref{nTitty}, the most general constitutive relations of the form \eqref{constitutiverelationszzzUEIT} are 
\begin{equation}\label{constitIsrStew}
 T^{\nu \rho} =\delta\indices{^\nu _\sigma} \delta\indices{^\rho _\lambda}  T^{\sigma \lambda} \, ,
 \quad 
 s^\nu = s^\nu(n^\sigma, T^{\sigma \lambda}) \, , 
  \quad 
  n^\nu = \delta\indices{^\nu _\sigma} \, n^\sigma \, .
\end{equation}
The only non-trivial constitutive relation that should be provided is the one for the entropy current, and this is where statistical mechanics is required. 

Concrete implementations of the Israel-Stewart theory are typically formulated directly in Lindblom's representation (see subsection \ref{Lindddblumz}): given the degrees of freedom in \eqref{nTitty},  a natural choice for the  dissipation fields is
\begin{equation}
\label{IlGufoStrapotente}
    (\mathbb{A}^A) = (\Pi,q^\sigma,\Pi^{\sigma \lambda}) \, ,
\end{equation}
where (introducing the projector $h_{\nu \rho}=g_{\nu \rho}+u_\nu u_\rho$)
\begin{equation}
\begin{split}
 & \Pi = \Delta T^{\alpha \beta} h_{\alpha \beta}/3 
\\
 & q^\sigma = -\Delta T^{\alpha \beta} h\indices{^\sigma _\alpha} u_\beta
\\
 & \Pi^{\sigma \lambda} = \Delta T^{\alpha \beta}  h\indices{^\sigma _\alpha}  h\indices{^\lambda _\beta} -\Pi h^{\sigma \lambda} \, , 
\end{split}
\end{equation}
can be interpreted as the bulk-viscous stress, the heat flux and the shear-viscous stress \cite{Liu1986}. 
Together, the fields in \eqref{IlGufoStrapotente} constitute the 9 additional degrees of freedom of the dissipative model, so that the constitutive relation for the entropy current can be equivalently rewritten in Lindblom's representation as 
\begin{equation}
    s^\nu =s^\nu (\phi^a,\mathbb{A}^A)= s^\nu (u^\sigma,n,\mathcal{U},\Pi,q^\sigma,\Pi^{\sigma \lambda}) \, .
\end{equation}
This formula is very convenient because it allows one to easily separate the perfect fluid part from the non-equilibrium correction, and possibly make a perturbative expansion in the latter.

To complete the model, one must specify also the hydrodynamic equations. It is natural to take the conservation laws \eqref{conservotutto} as hydrodynamic equations, producing 5 independent equations, but 9 additional equations are left unspecified. 
\citet{Liu1986} proposed an elegant technique for constructing the remaining 9 equations in such a way that all the principles of UEIT are respected. 
The idea is to postulate that there are two additional tensor fields $A^{\nu \rho \mu}$ and $I^{ \rho \mu}$, with constitutive relations
\begin{equation}\label{BBBNNNZZXX}
A^{\nu \rho \mu} = A^{\nu \rho \mu} (n^\sigma, T^{\sigma \lambda})   \quad \quad   I^{ \rho \mu} = I^{\rho \mu} (n^\sigma, T^{\sigma \lambda}) \, ,
\end{equation}
to be determined from microphysics (note the absence of gradients in the constitutive relations). Then, it is assumed that the remaining hydrodynamic equations of the model are  given by
\begin{equation}\label{AIMUNU}
\nabla_\nu A^{\nu \rho \mu} = I^{\rho \mu} \, .
\end{equation}
In this way, it is guaranteed that the final equations have the form \eqref{hydrouzZZUEIT}. However, equation \eqref{AIMUNU} contains in principle 16 independent equations, leading to an over-determination of the system. 
The solution proposed by \citet{Liu1986} is to require that
\begin{equation}\label{asxz}
A^{\nu \rho \mu} =A^{\nu (\rho \mu)}   \quad \quad   
I^{ \rho \mu} =I^{( \rho \mu)}
\end{equation}
and that (we have an additional - with respect to \citet{Liu1986} due to the opposite metric signature)
\begin{equation}\label{asxz2}
A\indices{^\nu ^\rho _\rho} = -m^2 \, n^\nu   \quad \quad   
I\indices{^\rho _\rho}=0 \, ,
\end{equation}
where $m$ is an appropriate constant. In this way, the independent equations are reduced from $16$ to $9$, closing the system. When the hydrodynamic equations of an Israel-Stewart theory can be presented in this form, we may say that the model is of \textit{divergence-type} \citep{rezzolla_book}.

Despite the construction of \citet{Liu1986} may seem just a sequence of assumptions, equations \eqref{BBBNNNZZXX}, \eqref{AIMUNU}, \eqref{asxz} and \eqref{asxz2} find their fundamental justification within kinetic theory \cite{cercignani_book}, where $A^{\nu \rho \mu}$ is the third moment of the particle distribution function and $I^{ \rho \mu}$ is its collision production (in which case $m$ is the particles' rest-mass). 
Interestingly, if $I^{ \rho \mu}$ is modelled consistently with kinetic theory, then the second law is directly enforced as a consequence of the H-theorem. 
However, it should be kept in mind that any UEIT model is fundamentally phenomenological, so that no particular kinetic interpretation needs to be provided for   $A^{\nu \rho \mu}$ and $I^{ \rho \mu}$ in principle: they can also be regarded just as elements of a prescription for constructing the hydrodynamic equations (in which case $m$ may be set conveniently to 0), as discussed with more details by \citet{GerochLindblom1990}.

In \citet{Israel_Stewart_1979} a less abstract way of deriving the hydrodynamic equations is proposed: a fiducial equilibrium state is taken, and the entropy current is expanded to the second order in the deviations from this state. Then, the constraints imposed by the second law (which in UEIT must be enforced as an exact relation at all the orders) are used to guess the hydrodynamic equations. The resulting model is what is commonly referred to as the original Israel-Stewart theory \citep{andersson2007review,rezzolla_book}. However, it is important to note that, depending on the choice of the fiducial equilibrium state, the equations that one is led to guess are different (this is just another manifestation of the problem of the hydrodynamic frames \cite{Kovtun2019}), so that  it is possible to construct many different Israel-Stewart theories \citep{Hishcock1983,OlsonLifsh1990}: the construction in terms of the phenomenological fields $A\indices{^\nu ^\rho ^\mu}$ and
$I\indices{^\rho ^\mu}$ directly reflects this freedom.

Most of the Israel-Stewart theories are known to be Lyapunov-Stable \citep{Hishcock1983,OlsonLifsh1990,GerochLindblom1990}, provided that the constitutive relations are realistic, in agreement with the general argument of subsection \ref{LyiYup}. Furthermore, if the stability conditions found by \citet{Hishcock1983} hold, then the entropy is a Lyapunov function of the system, at least close to the equilibrium state  \citep{GavassinoLyapunov_2020}.

\subsection{The Carter-Khalatnikov approach as a UEIT theory}

Carter's   approach to hydrodynamics finds its application in modelling reacting mixtures and conducting media (in particular heat conduction and superfluidity) in general relativity \citep{noto_rel}, and it is a convenient formalism for modelling neutron star interiors \citep{haskellsedrakian2017,chamel_super}.  Its constitutive relations can be equivalently derived either from a convective or from a potential variational principle \cite{Carter_starting_point}. 
A general introduction to Carter's approach may be found in  \citet{Carter_starting_point} and \citet{andersson2007review}, see also \citet{Termo} and references therein for more formal aspects.  

Contrarily to the premise \eqref{nTitty} of Israel and Stewart's model, now $T^{\mu \nu}$ is not treated  as a fundamental field. Instead, in the simplest formulation of his approach,  Carter postulates that it is possible to identify a set of currents $n_i^\sigma$ which constitute the degrees of freedom,
\begin{equation}\label{VARFNIU}
(\varphi_i) = (n_i^\sigma) \, .
\end{equation}
Two of the currents $n_i^\sigma$ can be chosen to be $s^\sigma$ and $n^\sigma$, while the remaining ones depend on the non-equilibrium thermodynamic properties of the system under consideration. This theory can thus be used to describe reacting mixtures, multi-temperature systems, or the viscous interaction between different chemical species\footnote{Although in the present work we make the simplifying assumption that there is only one conserved current, there may still be many non-conserved chemical species, such as photons.}. 

The most general constitutive relations of the form \eqref{constitutiverelationszzzUEIT}, that are compatible with the choice of fields \eqref{VARFNIU}, are
\begin{equation}\label{constitCARTKAL}
 T^{\nu \rho} = T^{\nu \rho} (n^\sigma_i)  \quad \quad 
 s^\nu = \delta\indices{^\nu _\sigma} \, s^\sigma \quad \quad 
  n^\nu = \delta\indices{^\nu _\sigma} \, n^\sigma \, .
\end{equation}
The only non-trivial constitutive relation one needs to determine is, therefore, the one for the stress-energy tensor. If, however, we assume that the 10 relations $T^{\nu \rho} = T^{\nu \rho} (n^\sigma_i)$ are all independent from each other, then we are left with the formidable task of computing, directly from microphysics, 10 different non-equilibrium ensemble averages separately. 
To avoid this problem, Carter postulates that all these 10 relations can be derived  from a single non-equilibrium equation of state:  there is a fundamental scalar field $\Lambda$ with a constitutive relation (up to now, the dependence of the fields on the metric was understood, but in this case it is important to keep track of it explicitly) 
\begin{equation}
\label{BendaSullOcchio}
\Lambda = \Lambda(n_i^\sigma, g_{\sigma \lambda}) \, ,
\end{equation}
that encodes, somehow, the non-equilibrium equation of state of the fluid\footnote{
The double role of $\Lambda$ as a thermodynamic potential and as a generating function for $T^{\nu \rho}$ is extensively discussed in \citep{Termo}.
}, and such that
\begin{equation}\label{TTcarter}
T^{\nu \rho} = \dfrac{2}{\sqrt{|g|}} \dfrac{\partial (\sqrt{|g|}\Lambda)}{\partial g_{\nu \rho}} \bigg|_{\sqrt{|g|} \, n_i^\sigma} \, .
\end{equation}    
The convenient feature of assuming the existence of $\Lambda$ and its constitutive relation \eqref{BendaSullOcchio} is that we can decompose $\nabla_\nu T\indices{^\nu _\rho}$ in a convenient way: for each current $n^\nu_h$, it is possible to define an associated covector 
\begin{equation}
\mu_\nu^h := \dfrac{\partial \Lambda}{\partial n_h^\nu} \bigg|_{n_i^\sigma,g_{\sigma \lambda}} \, 
\qquad (i,\sigma)\neq (h,\nu) \, ,
\end{equation}
so that the divergence of the stress-energy tensor \eqref{TTcarter} reads
\citep{Carter_starting_point}
\begin{equation}
\nabla_\nu T\indices{^\nu _\rho} = \sum_h \bigg( \mu^h_\rho \, \nabla_\nu n_h^\nu + 2n_h^\nu \, \nabla_{ [ \nu} \mu_{\rho ]}^h \bigg) \, .
\end{equation}
This suggests us that we may conveniently postulate that there are some covector fields $\mathfrak{R}^h_\rho$ (one for each independent current), with constitutive relations\footnote{
The more general constitutive relation  where there is also a dependence on the gradients of the currents, $\mathfrak{R}^h_\rho = \mathfrak{R}^h_\rho (n_i^\sigma,\nabla_\lambda n_i^\sigma)$, has also been considered for describing heat conduction \citep{Lopez09}, superfluid vortex dynamics \cite{langlois98,RauWasserman2020,GavassinoIordanskii2021} and radiation hydrodynamics \citep{GavassinoRadiazione}. }
\begin{equation}\label{emckfl}
\mathfrak{R}^h_\rho = \mathfrak{R}^h_\rho (n_i^\sigma) \, ,
\end{equation}
such that the hydrodynamic equations take the form
\begin{equation}\label{hydrunkz}
\mu^h_\rho \, \nabla_\nu n_h^\nu + 2n_h^\nu \, \nabla_{ [ \nu} \mu_{\rho ]}^h = \mathfrak{R}^h_\rho  \, . 
\end{equation}
These equations have the structure \eqref{hydrouzZZUEIT}. Furthermore, they allow us to convert the conditions \eqref{conservotutto} and \eqref{secondlaw} into algebraic constraints on the relations \eqref{emckfl}:
\begin{equation}
\sum_h \mathfrak{R}^h_\rho =0  \quad \quad  \mathfrak{R}^n_\rho n^\rho = 0 \quad \quad  \dfrac{\mathfrak{R}^s_\rho s^\rho}{\mu^s_\lambda s^\lambda} \geq 0 \, .
\end{equation} 
In conclusion, the theory meets by construction all the requirements for being a UEIT model.
 
Many extension of Carter's formalism have been proposed. For example, in an approach entirely based on currents it is not obvious how to account for the possible presence of shear viscosity. For this reason, \citet{carter1991} extended the approach to include, among the fields of the theory, some symmetric tensors $\tau_\Sigma^{\sigma \lambda}$, which contain the information about the shear viscosity contributions~\citep{PriouCOMPAR1991}.

Originally, Carter built his formalism by using a hybrid methodology: he derived the constitutive relation \eqref{TTcarter} from a convective variational approach, treating the scalar field $\Lambda$ as a Lagrangian density and postulating the hydrodynamic equations \eqref{hydrunkz} as natural fluid generalizations of Newton's second law \cite{carter1991,GavassinoRadiazione}. 
Given that the convective variational approach is particularly well-suited for \emph{multifluids} \cite{andersson2007review,Termo}, there is ongoing research on whether it is possible to derive the fully dissipative theory directly from a convective action principle. 
In fact, it has been recently noted  that a particular class of  theories resembling  Carter's original formalism can be derived directly from an action principle \citep{AnderssonComerDissAct2015,CeloraNils2020}. 
These theories are constructed within a Lagrangian specification of the flow field (which is the natural framework in which the convective variational approach is formulated \cite{andersson2007review}, see subsection \ref{perflabgeulftssa}), so that are not  presented in a natural UEIT form: the hydrodynamic equations for the fields $\Phi^i$ are of the second order and the constitutive relations involve first derivatives. 
It is yet not completely understood under which conditions  it is possible to perform a  change of variables and move to the Eulerian description. As a consequence, we do not know which conditions should be imposed on this kind of variational models to meet the formal requirements of UEIT. However, there are some specific cases in which such a mapping has been shown to be possible \citep{Andersson_relativisticplasmas}.

\section{A universal UEIT model for bulk viscosity}


As a concrete example of the usefulness of the UEIT point of view, we describe an almost\footnote{See, e.g.,  \citet{jain_kovtun_2020arXiv}.} universal model for bulk viscosity. In other words, we construct the most general bulk-viscous simple fluid model \citep{BulkGavassino} directly from the principles of UEIT, invoking almost no additional assumption. The resulting model encodes, as particular cases, the Israel-Stewart theory for bulk viscosity and Carter's approach for comoving species. 

\subsection{The fields of the theory}\label{axsummiz}


A fluid is purely bulk-viscous (no shear viscosity and no heat conduction) if, at each point of the spacetime, there exists an observer (defined by the four-velocity $u^\sigma$) for which the local fluid element is isotropic. 
Clearly, we can use $u^\sigma$ as a field of the theory. Thanks to the isotropicity assumption, we must have that
\begin{equation}
u^\sigma \,=\, 
{n^\sigma}/{\sqrt{-n^\lambda n_\lambda}} \,=\, 
{s^\sigma}/{\sqrt{-s^\lambda s_\lambda}} \, .
\end{equation}
The field $u^\sigma$ is the non-equilibrium notion of fluid velocity (no frame ambiguity for the fluid velocity can exist in this case).
We also construct the scalar fields $n$ and $\mathcal{U}$ using the prescriptions \eqref{proscrivuznz} and analogously
\begin{equation}
s=-s^\sigma u_\sigma.
\end{equation} 
By definition, the fields $n$, $\mathcal{U}$ and $s$ are the densities of  particles, energy and entropy, as measured by a local observer moving with $u^\sigma$. Since, in this case, it is more convenient to work with quantities per particle, we define
\begin{equation}
v \,=\, {1}/{n}  
\quad \quad \quad    
\tilde{\mathcal{U}} \,=\,  {\mathcal{U}}/{n}  
\quad \quad \quad  
x_s \,=\,  {s}/{n} \, ,
\end{equation}
which are the volume per-particle, the energy per-particle, and the entropy per-particle (i.e. the entropy fraction), as measured by the local observer moving with $u^\sigma$.

Since the vector field $u^\sigma$ uniquely identifies the only preferred direction defined by the fluid motion, all the remaining fields $\varphi_i$ of the theory can be taken to be scalar fields:
\begin{equation}\label{DoFbelli}
(\varphi_i) = (u^\sigma,v,x_s,\alpha_1,...,\alpha_{\mathfrak{D}-5}) \, ,
\end{equation}
where $\alpha_1,...\alpha_{\mathfrak{D}-5}$ are additional $\mathfrak{D}-5$ independent scalar fields\footnote{
Contrarily to Lindblom's dissipation fields $\mathbb{A}^A$, the  $\alpha_A$ do not need to vanish in local thermodynamic equilibrium. On the other hand, we will soon introduce some alternative variables which can be treated as dissipation fields.
}. 
Their number, properties and physical meaning are dictated only by microphysics and can change from fluid to fluid. 

The symmetry prescriptions, combined with the principles of UEIT, reduce the space of the allowed constitutive relations to
\begin{equation}\label{iobulko}
\begin{split}
 & T^{\nu \rho} = (n \tilde{\mathcal{U}} +\Psi)u^\nu u^\rho + \Psi g^{\nu \rho}
 \\ 
 & s^\nu = x_s nu^\nu 
 \\ 
 & n^\nu = n u^\nu \, ,
\end{split}
\end{equation}
with
\begin{equation}\label{quiebulk}
\tilde{\mathcal{U}}=\tilde{\mathcal{U}}(v,x_s,\alpha_A)   \quad \quad   \Psi=\Psi(v,x_s,\alpha_A) \, .
\end{equation} 
We have introduced the abstract index $A=1,...,\mathfrak{D}-5$ for later notation convenience (Einstein's convention for this index will be applied). Equation \eqref{quiebulk} shows us that the fields $\alpha_A$ can be interpreted as the additional non-conserved (i.e. of type $\xi_i$, see subsection \ref{basilKs}) thermodynamic variables of an extended non-equilibrium equation of state. 
The non-equilibrium generalizations of the thermodynamic pressure and temperature may be defined as
\begin{equation}\label{thermoduzzo}
P := - \, \dfrac{\partial \tilde{\mathcal{U}}}{\partial v} \bigg|_{x_s,\alpha_A}   \quad \quad \quad   
\Theta := \dfrac{\partial \tilde{\mathcal{U}}}{\partial x_s} \bigg|_{v,\alpha_A} \, ,   
\end{equation}
however, it is important to keep in mind that, out of equilibrium, a unique proper notion of pressure and temperature does not exist. One should not be tempted to attribute to $P$ and $\Theta$ a primary physical relevance. 
In fact, it is always possible to perform a field redefinition, introducing arbitrary new fields $\alpha'_B = \alpha'_B (v,x_s,\alpha_A)$, and to take the partial derivatives in \eqref{thermoduzzo} at constant $\alpha'_B$ rather than at constant $\alpha_B$, obtaining a different result. 
As a consequence, out of equilibrium we have that in general $P \neq \Psi$, since $\Psi$ is the physical stress tensor, which is necessarily invariant under field-redefinitions. 
Finally, we introduce the generalised affinities
\begin{equation}\label{AQAQA}
\mathbb{A}^A := - \, \dfrac{\partial \tilde{\mathcal{U}}}{\partial \alpha_A} \bigg|_{v,x_s},
\end{equation}
so we can write the differential of the energy per-particle explicitly as
\begin{equation}\label{differenz}
d\tilde{\mathcal{U}} \, = \, -P dv + \Theta dx_s -\mathbb{A}^A d\alpha_A \, . 
\end{equation} 
The interpretation of $\mathbb{A}^A$ as generalised affinities \cite{PrigoginebookModernThermodynamics2014} comes from the fact that, for given $v$ and $\tilde{\mathcal{U}}$, the entropy per particle $x_s$ is maximised when
\begin{equation}\label{AAAAA}
\mathbb{A}^A=0   \quad \quad   \forall A \, .
\end{equation}
Clearly, this is the condition of local thermodynamic equilibrium, allowing us to identify $\mathbb{A}^A$ with the Lindblom dissipation fields, see subsection \ref{Lindddblumz}.

\subsection{Hydrodynamic equations for pure bulk-viscous fluids}

Let us, now, move to the dynamical equations of the system.
The conservation laws \eqref{conservotutto} can be written in the more convenient form
\begin{equation}\label{bulkino}
\begin{split}
& (n\tilde{\mathcal{U}} + \Psi) u^\rho \nabla_\rho u_\nu =- (\delta\indices{^\rho _\nu} + u^\rho u_\nu)\nabla_\rho \Psi   \\
& \dot{\tilde{\mathcal{U}}} =-\Psi \dot{v} \\
& \dot{v}=v \nabla_\nu u^\nu ,\\
\end{split}
\end{equation}
where we have introduced the notation $\dot{f}:= u^\nu \nabla_\nu f$ for any tensor field $f$. The second equation can be equivalently rewritten as
\begin{equation}\label{entropyIo}
\Theta \dot{x}_s = (P-\Psi)\dot{v} + \mathbb{A}^A \dot{\alpha}_A,
\end{equation} 
where we made use of the differential \eqref{differenz}.

Given that the degrees of freedom of the theory are the fields $\varphi_i$ themselves, we know from equation \eqref{DoFbelli} that we need $\mathfrak{D}$ first-order equations to close the system. However, equation \eqref{bulkino} already provides  5 equations, so we need only other $\mathfrak{D}-5$ equations. Recalling the symmetry prescriptions, these equations can be presented in the form
\begin{equation}\label{qowmsk}
\mathfrak{F}_A(\varphi_i , \nabla_\sigma \varphi_i)=0   \quad \quad    A=1,...,\mathfrak{D}-5 \, ,
\end{equation}
where each of the $\mathfrak{F}_A$ is a scalar. 
Now, if we go to the local inertial frame of an observer instantaneously comoving with $u^\sigma$, we can rewrite \eqref{qowmsk} as
\begin{equation}\label{ewdkomc}
\mathfrak{F}_A(v,x_s,\alpha_B , \dot{v},\dot{x}_s,\dot{\alpha}_B, \dot{u}^j, \partial_j \varphi_i)=0 \, ,
\end{equation}
where we have used the fact that $\dot{u}^0=0$ in this reference frame. 
Given that the fluid element is isotropic in this reference frame (and that $\partial_j u^j = \dot{v}/v$), the only way for the scalars $\mathfrak{F}_A$ to have a dependence on $\dot{u}^j$ and $\partial_j \varphi_i$ is through quadratic couplings of the kind
\begin{equation}
\dot{u}^j \dot{u}_j   \quad \quad    \dot{u}^j \partial_j \varphi_i   \quad \quad   \partial^j \varphi_i \partial_j \varphi_{h}.
\end{equation}
Although the presence of such terms may not be excluded a priori, we expect them to become relevant only for very large gradients, and we will neglect them. 
In addition, we can use equation \eqref{entropyIo} to replace $\dot{x}_s$ inside \eqref{ewdkomc}, to eliminate the explicit dependence of $\mathfrak{F}_A$ on $\dot{x}_s$. Therefore, we are left with 
\begin{equation}\label{ewdkomc33}
\mathfrak{F}_A(v,x_s,\alpha_B , \dot{v},\dot{\alpha}_B)=0 \, .
\end{equation}
Finally, we note that these equations can be interpreted as a system of $\mathfrak{D}-5$ algebraic equations for the $\mathfrak{D}-5$ unknowns $\dot{\alpha}_B$.  We can imagine to solve them and finally obtain
\begin{equation}\label{hybbo}
\dot{\alpha}_A = f_A (v,x_s,\alpha_B , \dot{v}) \, .
\end{equation} 
The combination of \eqref{hybbo} with \eqref{bulkino} constitutes the complete system of $\mathfrak{D}$ first-order differential equations governing the evolution of a generic bulk-viscous fluid within UEIT.

\subsection{The structure of the universal model}\label{jejnjujskhahm}

There is a final step we can make to strongly simplify the structure of the universal model. Let us expand \eqref{hybbo} at the first order in $\dot{v}$:
\begin{equation}\label{faf}
\dot{\alpha}_A  = D_A  + \dot{v} \, \mathcal{K}_A ,
\end{equation}
where $D_A$ and $ \mathcal{K}_A $ are functions of $(v,x_s,\alpha_B)$ only. It may seem that, by performing this expansion, we are restricting the validity of the model only to slow systems. However, it should be kept in mind that ``small $\dot{v}$'' is not necessarily synonym of ``slow evolution'', since Thermo-modes can occur also in static fluids (namely when $u^\sigma = \delta\indices{^\sigma _t}=\text{const}$). Furthermore, in \cite{BulkGavassino} it is shown that \eqref{faf} happens to be an almost exact relation for most physical systems (in any regime in which a UEIT description is applicable).

Inserting \eqref{faf} into \eqref{entropyIo} and recalling \eqref{conservotutto} and \eqref{secondlaw} we obtain
\begin{equation}\label{intermadioAAAA}
v\Theta \nabla_\nu s^\nu= 
\big(P+\mathbb{A}^A \mathcal{K}_A -\Psi \big)\dot{v} + \mathbb{A}^A D_A \geq 0 \, ,
\end{equation}
where we have assumed $\Theta \geq  0$.\footnote{this is certainly the case at equilibrium, when $\Theta$ acquires the proper meaning of temperature.}
Now, let us work in the limit in which the dissipation fields $\mathbb{A}^A$ are small (i.e. close to local thermodynamic equilibrium) and consider, first, the case in which $\dot{v}=0$. If we want the second law to be respected, we need to impose $\mathbb{A}^A D_A \geq 0$ for any small $\mathbb{A}^A$. This implies that $D_A$ must be at most linear in $\mathbb{A}^A$ at the leading order: 
\begin{equation}
D_A \approx v \, \Xi_{AB}\mathbb{A}^B  \quad   \text{with} \quad   \mathbb{A}^A \Xi_{AB} \mathbb{A}^B \geq 0 \qquad   \forall \, \mathbb{A}^A  .  
\end{equation}
Now, let us move to the case in which $\dot{v}$ has a finite value. Since $\mathbb{A}^A D_A=v \,\mathbb{A}^A \Xi_{AB} \mathbb{A}^B$ is an order 2 in the dissipation fields, the term $\big(P+\mathbb{A}^A \mathcal{K}_A -\Psi \big)\dot{v}$ needs to be in turn an order $(\mathbb{A}^A)^2$, to ensure the validity of the second law, implying that $P+\mathbb{A}^A \mathcal{K}_A -\Psi =0$ to the first order in $\mathbb{A}^A$. Thus we have shown that $\Psi$ and $P$ must be related, in a way that depends on the coefficients $\mathcal{K}_A$, which contain information about the intrinsic evolution of the fields $\alpha_A$ under a volume expansion. An analogous identity has been derived on thermodynamic grounds in \cite{BulkGavassino}, where it is also shown that such a relation, which has been proved here for small dissipation fields, should hold also far from equilibrium (provided that a UEIT description remains possible). 

In conclusion, we are allowed to split \eqref{intermadioAAAA} into the two following separate relations:
\begin{equation}\label{LURRO}
\Psi=P+\mathbb{A}^A \mathcal{K}_A   \quad \quad  \Theta \dot{x}_s =  \mathbb{A}^A D_A \approx v \, \mathbb{A}^A \Xi_{AB}  \mathbb{A}^B,
\end{equation} 
where $\approx$ is valid only close to equilibrium, while $=$ is assumed to always hold.
Note that in equilibrium the first relation becomes $\Psi=P$, so that the thermodynamic pressure becomes the isotropic stress.
It is also useful to remember that we can always rewrite the second equation of \eqref{LURRO} in the equivalent form
\begin{equation}\label{eselaproducoio}
\Theta \nabla_\nu s^\nu = n \mathbb{A}^A D_A \geq 0 \, .
\end{equation}

To summarize, given a choice of fields $\alpha_A$, the ingredients that need to be computed from microphysics to build a UEIT model for bulk viscosity are $2\mathfrak{D}-9$ constitutive relations, namely an equation of state for $\tilde{\mathcal{U}}$ and the formulas for the $2(\mathfrak{D}-5)$ kinetic coefficients $D_A$ and $\mathcal{K}_A$.

\subsection{Invariance under field redefinitions}\label{fiedrif}

As we anticipated, there are infinitely many valid choices for the fields $\alpha_A$ in a UEIT model for bulk viscosity. Therefore, it is important to verify that the equations of the theory are invariant under a field redefinition. 
To check this invariance property, we perform a generic change of variables
\begin{equation}\label{fieldrefegrf}
(u^\sigma,v, x_s, \alpha_A) \longrightarrow (u^\sigma,v, x_s, \alpha'_B) \, ,
\end{equation}
where $\alpha'_B = \alpha'_B (v,x_s,\alpha_A)$ are $\mathfrak{D}-5$ new arbitrary independent scalar fields. Rewriting the differential \eqref{differenz} in terms of $\alpha'_B$ we find
\begin{equation}
d\tilde{\mathcal{U}} =-P' dv + \Theta' dx_s -\mathbb{B}^B d\alpha'_B \, , 
\end{equation} 
with
\begin{equation}\label{iPrimati}
 \begin{split}
& P=P'+ \mathbb{B}^B \dfrac{\partial \alpha_B'}{\partial v}  
 \\
& \Theta = \Theta' - \mathbb{B}^B \dfrac{\partial \alpha_B'}{\partial x_s} 
 \\  
 & \mathbb{A}^A = \mathbb{B}^B \dfrac{\partial \alpha_B'}{\partial \alpha_A} \, .  
\end{split}
\end{equation}
The hydrodynamic equations of the new fields can be obtained from the chain rule:
\begin{equation}
\dot{\alpha}'_B = \dfrac{\partial \alpha_B'}{\partial v} \dot{v} + \dfrac{\partial \alpha_B'}{\partial x_s} \dot{x}_s + \dfrac{\partial \alpha_B'}{\partial \alpha_A} \dot{\alpha}_A  
\end{equation}
and it is immediate to verify that they have exactly the same form as \eqref{faf},
\begin{equation}\label{dprim}
\dot{\alpha}'_B = D'_B + \dot{v} \mathcal{K}'_B,
\end{equation}
with
\begin{equation}\label{reuommi}
 D'_B = \dfrac{\partial \alpha_B'}{\partial \alpha_A} D_A + \dfrac{\partial \alpha'_B}{\partial x_s} \dfrac{\mathbb{A^A }D_A}{\Theta} \, ,
 \quad  
 \mathcal{K}'_B = \dfrac{\partial \alpha_B'}{\partial \alpha_A} \mathcal{K}_A + \dfrac{\partial \alpha'_B}{\partial v} \, .
\end{equation}
Now, using equations \eqref{iPrimati}, \eqref{LURRO} and \eqref{reuommi}, it is finally possible to prove the full covariance of the theory under field redefinitions, namely
\begin{equation}
\Psi=P'+\mathbb{B}^B \mathcal{K}'_B  \quad \quad   \Theta' \dot{x}_s =  \mathbb{B}^B D'_B \, .
\end{equation}  
This shows that, since $\Psi$ is invariant under field redefinitions, but $P$ is not, the piece $\mathbb{A}^A \mathcal{K}_A$ in \eqref{LURRO} plays the role of a counter-term which restores formal covariance, fixing the intuitive (but non-covariant) relation ``$\Psi = P$''.\footnote{The idea of including additional terms which restore the formal covariance of a theory under field redefinitions is a central feature of field theory and has important implications in UEIT. In Carter's approach, for example, this is known as the problem of the \emph{chemical basis} \cite{Carter_starting_point,GavassinoRadiazione}, which in neutron star crusts results into a \emph{chemical gauge} freedom \cite{carter_macro_2006}, which can have serious implications for superfluid vortex dynamics \cite{GavassinoIordanskii2021}.}

\subsection{Obtaining a reacting mixture using a field redefinition}\label{chemistriuz}

The formal covariance of the theory under field redefinitions of the kind \eqref{fieldrefegrf} can be exploited to show that any theory for bulk viscosity, when modelled within UEIT, can be mapped into an effective chemical mixture of comoving species, if a convenient choice of fields is adopted \citep{BulkGavassino}. 
To show this, we start with arbitrary fields $\alpha_A$ and then perform a field redefinition, introducing the $\mathfrak{D}-5$ effective fractions $x_B = x_B(v,x_s,\alpha_A)$, such that
\begin{equation}
\mathcal{K}'_B = \dfrac{\partial x_B}{\partial \alpha_A} \, \mathcal{K}_A + \dfrac{\partial x_B}{\partial v} =0 \, .
\end{equation}
This change of variables $\alpha_A \rightarrow x_A$ is always possible, and it provides a way to straighten the vector field 
\begin{equation}
W_{\text{fr}} \, := \,  
\mathcal{K}_A \, \dfrac{\partial}{\partial \alpha_A}\bigg|_{v,x_s} + \dfrac{\partial}{\partial v} \bigg|_{x_s,\alpha_A}
 = \,  \dfrac{\partial}{\partial v} \bigg|_{x_s,x_B} ,
\end{equation} 
which is the generator of adiabatic expansions (i.e. reversible changes in $v$) over the manifold of the thermodynamic states $\{(v,x_s,\alpha_A)\}$: this is the reason why it is convenient to to make the choice of fields $(u^\sigma,v,x_s,x_A)$, see Section 2 of \citep{BulkGavassino}. 
Thanks to these new variables, we have that
\begin{equation}
\Psi = P' = - \, \dfrac{\partial \tilde{\mathcal{U}}}{\partial v} \bigg|_{x_s,x_B}  ,
\end{equation}
which is consistent with the interpretation of $x_B$ as effective chemical fractions, or reaction coordinates \citep{prigoginebook}. The equations of motion \eqref{dprim} reduce to
\begin{equation}
\dot{x}_B = D_B' \, .
\end{equation}
If we define the effective chemical currents
\begin{equation}
n_B^\nu := x_B n^\nu \, ,
\end{equation}
we obtain the familiar equation
\begin{equation}
\nabla_\nu n_B^\nu = nD'_B \approx \Xi_{BC}' \mathbb{B}^C \, ,
\end{equation}
which shows us that we can interpret $nD'_B$ as some effective reaction rates. In fact, the entropy production reads
\begin{equation}
\Theta' \nabla_\nu s^\nu = n \, \mathbb{B}^B D'_B \approx \mathbb{B}^B \Xi_{BC}'\mathbb{B}^C \geq 0 \, , 
\end{equation}
which shows us that in the non-dissipative limit $\nabla_\nu n_B^\nu =0$. The description that we have obtained is fully equivalent to the one of a reacting mixture \cite{noto_rel}, showing that, within UEIT, \emph{every model for pure bulk viscosity has a chemical analogue}, whatever the mechanism responsible for dissipation. It also implies that multi-temperature fluids, reacting mixtures and perfect-fluid radiation-hydrodynamics (in the limit of infinite elastic scattering opacity) are, from the point of view of UEIT, nothing more than bulk-viscous fluids, where any non-equilibrium degree of freedom is  a source for bulk viscosity \citep{GavassinoRadiazione}.

\subsection{Israel-Stewart as a particular case of the universal model}

The Israel-Stewart theory for bulk viscosity is a UEIT model with $\mathfrak{D}=6$, where the bulk-viscous part $\Pi$ of the isotropic stress is the only additional dynamical variable of the kind introduced in \eqref{DoFbelli}, namely $\alpha := \Pi$, so that the fields of the theory are $(u^\sigma,v, x_s,\Pi)$.

To simplify the calculations we work close to thermodynamic equilibrium, where we can assume that the hydrodynamic equation for $\Pi$ can be approximated by the ``truncated equation'' \citep{Salmonson1991,Zakari1993,Maartens1995}
\begin{equation}
\label{JonasSalmon}
\tau \dot{\Pi} + \Pi = -\zeta \nabla_\nu u^\nu \, , 
\end{equation} 
where the viscosity coefficient $\zeta(v, x_s)>0$ determines the magnitude of the isotropic stress relative to expansion in the limit $\tau\rightarrow0$, where the usual (non-dynamical) Navier-Stokes relation $\Pi = -\zeta \nabla_\nu u^\nu$ is recovered.
Arranging the telegraph-type equation \eqref{JonasSalmon} to match the standard form in \eqref{faf}, gives
\begin{equation}
\dot{\Pi} = - \, \dfrac{\Pi}{\tau} - \dfrac{\zeta \dot{v}}{\tau v} 
 \, \, \Rightarrow   \, \,  
D_\Pi = - \, \dfrac{\Pi}{\tau} \, , \quad 
\mathcal{K}_\Pi = - \dfrac{\zeta }{\tau v} \, .
\end{equation}
\citet{Israel_Stewart_1979} assumed an energy per-particle expanded up to the second order in $\Pi$, namely
\begin{equation}
\tilde{\mathcal{U}}(v,x_s,\Pi)= \tilde{\mathcal{U}}_{\text{eq}}(v,x_s) + \dfrac{\beta_0(v,x_s) }{2n} \, \Pi^2 \, ,
\end{equation}
where $\beta_0 > 0$ to implement the minimum energy principle \cite{Callen_book}.
From \eqref{thermoduzzo} and \eqref{AQAQA}, we find that
\begin{equation}
P\approx - \, \dfrac{\partial \tilde{\mathcal{U}}_{\text{eq}}}{\partial v}  
\quad \quad  
\Theta \approx - \, \dfrac{\partial \tilde{\mathcal{U}}_{\text{eq}}}{\partial x_s}  
\quad \quad   
\mathbb{A}^\Pi = - \, \dfrac{\beta_0 \Pi}{n} \, ,
\end{equation}
where we have made a first-order truncation in $\Pi$ (it is evident that in this theory both $\Pi$ and $\mathbb{A}^\Pi$ can be considered Lindblom's dissipation fields, as they both vanish in equilibrium). Imposing the defining relation for $\Pi$,
\begin{equation}
\Psi = - \, \dfrac{\partial \tilde{\mathcal{U}}_{\text{eq}}}{\partial v}  +\Pi \approx P+\Pi \, ,
\end{equation}
and requiring its consistency with the first identity of \eqref{LURRO}, we immediately find the formula of Israel-Stewart for the Thermo-mode relaxation time-scale \cite{rezzolla_book}:
\begin{equation}
\label{JonasSmargiasso}
\tau = \beta_0  \zeta >0 \, ,
\end{equation}
showing that $\tau$, $\beta_0$ and $\zeta$ are not independent coefficients. This condition can be used to show that equation \eqref{eselaproducoio} for the entropy production reduces to
\begin{equation}\label{quandoIsrproduceentrop}
\Theta \nabla_\nu s^\nu = {\Pi^2} \, / \, \zeta \, ,
\end{equation}
which is consistent with the Israel-Stewart theory \citep{Hishcock1983}. 

\section{A simple UEIT application: radiation-mediated bulk viscosity}
\label{TwoTemper}

We conclude our survey with a  direct application of the UEIT formalism. We derive the formula of \citet{Weinberg1971} for the bulk viscosity of a perfect fluid coupled with photon radiation directly from non-equilibrium thermodynamics, by using only the tools of UEIT. The original derivation of Weinberg was based on the approximate (slow-limit based) solution of the Boltzmann equation for the photon gas found by \citet{Thomas1930}. On the contrary, in our approach, no reference to a slow limit is required and there is no need to solve the kinetic equation explicitly.


\subsection{The Udey-Israel argument} 

Building on the kinetic calculations of \citet{Thomas1930}, \citet{Weinberg1971} showed that the dissipative interaction between a matter fluid (with short mean free path) and a radiation gas (with a much longer mean free path) can give rise to an effective radiation-mediated bulk viscosity. Later, \citet{UdeyIsrael1982} proposed a simple and intuitive interpretation for this phenomenon. They considered that, when a fluid element undergoes an adiabatic expansion  over a time-scale shorter than the one defined by the opacity (i.e. $\sim 1/\chi$, see subsection \ref{JonasCurioso}), then matter and radiation do not have time to interact. This implies that the two components will expand independently, each of them following its own adiabatic curve. This generates (if matter is not an ultra-relativistic ideal gas) a temperature difference between the two components. The consequent exchange of heat, which tries to re-equalise the two temperatures, gives rise to dissipation.

By using the methods of the so-called $M_1$ closure scheme (see e.g. \citet{Sadowski2013}), it has been recently verified that radiation-mediated bulk viscosity can be, indeed, modelled as a pure two-temperature effect \citep{GavassinoRadiazione}: this provides an explicit realization of the intuition of Udey and Israel. Here, we revise this argument  from the point of view of UEIT, proving also its complete consistency with the kinetic derivation of \citet{Weinberg1971}. Our main goal is to convince the reader that UEIT (like EIT) is not just ``hyperbolic Navier-Stokes'', but it is really a branch of non-equilibrium thermodynamics \cite{AnilePavon1998}.

\subsection{Fluid+radiation system: constitutive relations and hydrodynamic equations}

Following the same logic outlined in subsection \ref{axsummiz}, we isolate the phenomenon of bulk viscosity and remove shear viscosity and heat conduction by imposing an assumption of local isotropy for the  local observers defined by the four-velocity field $u^\sigma$. In addition, we beforehand   assume that it is enough to model the fluid+radiation system as a two-temperature fluid, in which we allow the radiation gas to have a different temperature with respect to the one of the matter fluid \citep{UdeyIsrael1982}. For simplicity, the matter fluid is treated as a perfect fluid (after all  we want to isolate and study the pure effects induced by the presence of radiation). 
Hence, we choose the fields of the  model as
\begin{equation}
\label{JonasKobras}
(\varphi_i) \,  = \, (u^\sigma,s_n,n,\Theta_\gamma) \,  ,
\end{equation}
where $s_n$ and $n$ are respectively entropy and particle density of the matter fluid (as measured in the frame of $u^\sigma$), while $\Theta_\gamma$ is the temperature of the radiation fluid. 
Given \eqref{JonasKobras}, the resulting UEIT model will necessarily have $\mathfrak{D}=6$ degrees of freedom. By symmetry, the constitutive relations are given by
\begin{equation}
\label{RRad}
\begin{split}
 & T^{\nu \rho} = (\mathcal{U} +\Psi)u^\nu u^\rho + \Psi g^{\nu \rho} \\
 & s^\nu = s u^\nu \\
 & n^\nu = n u^\nu  \, .
\end{split}
\end{equation}
Recalling that we assumed a perfect fluid model for the matter sector, we impose the equation of state of the full matter+radiation system to be separable into a matter and a radiation part  (see e.g. \citep{mihalas_book,rezzolla_book}):
\begin{equation}\label{Uradiuz}
\mathcal{U} = \rho(s_n,n) + a_R \Theta_\gamma^4 \, ,
\end{equation}
where $\rho$ is the energy density of the matter fluid (as measured in the frame of $u^\sigma$), so that 
the temperature $\Theta_n$ and the chemical potential $\mu$ of the matter fluid can be computed from the differential
\begin{equation}
d\rho = \Theta_n ds_n + \mu dn \, .
\end{equation}
We assume that also the total isotropic stress $\Psi$ can be decomposed:
\begin{equation}\label{Psiradiuz}
\Psi = P_n + \dfrac{1}{3} a_R \Theta_\gamma^4 \, ,
\end{equation}
where
\begin{equation}
\rho + P_n = \Theta_n s_n + \mu n \, .
\end{equation}
Similarly, the total entropy density is the sum of the matter and the radiation contributions \cite{Groot1980RelativisticKT},
\begin{equation}\label{Sradiuz}
s= s_n + \dfrac{4}{3} a_R \Theta_\gamma^3 \, .
\end{equation}
Now that we have the explicit definitions for all the symbols appearing into the constitutive relations \eqref{RRad}, we only need to assign the hydrodynamic equations. The conservation laws in \eqref{conservotutto} already provide 5 hydrodynamic equations. The missing equation can be computed directly from kinetic arguments. In fact, under the aforementioned assumptions, equation \eqref{RaZZAZA} implies (for small $\Theta_n-\Theta_\gamma$)
\begin{equation}\label{thetagamma}
\dot{\Theta}_\gamma = 
\chi \, (\Theta_n-\Theta_\gamma) - \dfrac{\Theta_\gamma}{3} \,  \nabla_\nu u^\nu \, ,
\end{equation}
which is consistent with the intuition of \citet{UdeyIsrael1982}. In fact, during and expansion, in the absence of matter-radiation interactions (i.e. when $\chi=0$), the radiation gas evolves along its own adiabatic curve (namely, $\Theta_\gamma^3 v = \text{const}$ \cite{Leff2002}).

\subsection{The fluid+radiation system as a UEIT model for bulk viscosity}

If we make the change of variables $(u^\sigma,s_n,n,\Theta_\gamma) \longrightarrow (u^\sigma,v,x_s,\alpha)$, then the fields of our system are exactly those of a UEIT theory for bulk viscosity with $\mathfrak{D}=6$. More explicitly, the change of variables is  
\begin{equation}
x_s = s(s_n,\Theta_\gamma)/n  \quad \quad 
  v = 1/n \quad \quad  
\alpha = \Theta_\gamma \, , 
\end{equation}
where the total entropy $s(s_n,\Theta_\gamma)$ is given in \eqref{Sradiuz}. Now, the constitutive relations \eqref{RRad} already have the form \eqref{iobulko}. Furthermore, the hydrodynamic equation \eqref{thetagamma} has the structure \eqref{faf}, with
\begin{equation}
D = \chi (\Theta_n-\Theta_\gamma)   \quad \quad   \mathcal{K}=- \, \dfrac{1}{3} n \Theta_\gamma \, .
\end{equation}
Hence, this simple hydrodynamic model is a legit UEIT model for bulk viscosity, provided that the second law is valid for any value of $\dot{v}$. The validity of the second law is guaranteed if the conditions \eqref{LURRO} are respected and $\mathbb{A}D \geq 0$. To verify this, we first need to compute the thermodynamic derivatives given in \eqref{differenz}:
\begin{equation}
\begin{split}
& P = - \, \dfrac{\partial}{\partial v} \bigg( \dfrac{\mathcal{U}}{n} \bigg) \bigg|_{x_s,\Theta_\gamma} = P_n + a_R \Theta_\gamma^3 \bigg( \dfrac{4}{3} \Theta_n -\Theta_\gamma \bigg) \\
& \Theta = \dfrac{\partial}{\partial x_s} \bigg( \dfrac{\mathcal{U}}{n} \bigg) \bigg|_{v,\Theta_\gamma} = \Theta_n \\
& \mathbb{A}=- \, \dfrac{\partial }{\partial \Theta_\gamma}\bigg( \dfrac{\mathcal{U}}{n} \bigg) \bigg|_{v,x_s} = \dfrac{4a_R \Theta_\gamma^2}{n} (\Theta_n - \Theta_\gamma) \, ,
\end{split}
\end{equation}
that can be easily used  to obtain a formula for the isotropic stress, 
\begin{equation}
\Psi  \,= \, P_n + \dfrac{1}{3} a_R \Theta_\gamma^4 \, = \,P + \mathbb{A} \mathcal{K} \, .
\end{equation}
Note that this relation holds arbitrarily far from equilibrium (i.e. for large values of $\Theta_n-\Theta_\gamma$), in accordance with the general discussion of subsection \ref{jejnjujskhahm}.

Finally, combining the second and third equation of \eqref{bulkino} with the constitutive relations \eqref{Uradiuz}, \eqref{Psiradiuz} and \eqref{Sradiuz}, thanks to  the hydrodynamic equation \eqref{thetagamma} we can compute the entropy production,
\begin{equation}
\label{quandoiorpprodent}
\dfrac{\Theta \nabla_\nu s^\nu}{n} = \Theta_n \, \dot{x}_s = \chi \dfrac{4 a_R \Theta_\gamma^2}{n} (\Theta_n - \Theta_\gamma)^2 = \mathbb{A}D \geq 0 \, ,
\end{equation}
which is in agreement with the second condition in \eqref{LURRO}. 
This completes the proof  that the matter+radiation fluid is a UEIT bulk-viscous fluid.

In subsection \ref{chemistriuz} we showed that there is always a change of variables which converts a UEIT model into an effective description for a reacting mixture. To find this description one has to make a change of variables that allows to straighten the vector field 
\begin{equation}
W_{\text{fr}}=  - \, \dfrac{\Theta_\gamma}{3v}\dfrac{\partial}{\partial \Theta_\gamma} \bigg|_{v,x_s} 
+ \dfrac{\partial}{\partial v} \bigg|_{x_s,\Theta_\gamma} \, .
\end{equation}
The simplest variable that is conserved along the flux generated by $W_{\text{fr}}$ is
\begin{equation}\label{xsgMama}
x_{s\gamma} \, = \, {4 \, a_R \, \Theta_\gamma^3 \, v}\, / \,3 .
\end{equation} 
Under this field redefinition, the theory reduces to the hydrodynamic model for radiation-mediated bulk viscosity presented in \citet{GavassinoRadiazione} (neglecting the photon-number effects).

\subsection{Recovering Weinberg's formula for the bulk viscosity coefficient}

The formalism of UEIT allows us to find the formula for the radiation-mediated bulk viscosity coefficient $\zeta$ by matching the UEIT model with Israel-Stewart  for small deviations from equilibrium, something that can always be done since both models have $\mathfrak{D}=6$ \citep{BulkGavassino}.

The first step consists of computing the viscous stress $\Pi$, which is defined as the first-order difference between the physical stress $\Psi$ and the stress that would be generated by the fluid if it were in local thermodynamic equilibrium with the same $n$ and $\mathcal{U}$. We adopt the following notation: given a thermodynamic variable $f$, we call $f$ its physical value and $f+\delta f$ the value that $f$ would have if the fluid were in local thermodynamic equilibrium with the same $n$ and $\mathcal{U}$. 
Hence, we can impose
\begin{equation}\label{nups}
\delta \Psi = -\Pi   \quad \quad\quad  
\delta n =0  \quad \quad \quad  
\delta \mathcal{U}=0 \, .
\end{equation}
In addition, from the minimum energy principle, we know that the affinity $\mathbb{A}$ must vanish at equilibrium, so we must require that
\begin{equation}
\mathbb{A}+\delta \mathbb{A}=0 \, .
\end{equation} 
Combining this condition with the second and the third equations of \eqref{nups} we obtain, working at the first order in the differences $\delta f$,
\begin{equation}\label{expendo}
\begin{split}
& \delta \Theta_\gamma = \dfrac{c_v^n}{c_v^n + c_v^\gamma} (\Theta_n - \Theta_\gamma)   
\\
& \delta \Theta_n =- \dfrac{c_v^\gamma}{c_v^n + c_v^\gamma} (\Theta_n - \Theta_\gamma) \, , 
\end{split}
\end{equation}
where  the heat capacities (per unit volume, at constant volume) are given by
\begin{equation}
c_v^n = \dfrac{\partial \rho}{\partial \Theta_n} \bigg|_{n}   
\qquad  
c_v^\gamma = \dfrac{d (\,a_R\, \Theta_\gamma^4\,)}{d \Theta_\gamma} = 4 \, a_R \, \Theta_\gamma^3 \, .
\end{equation}
Note that equation \eqref{expendo} is just the formula for the variation of the temperatures of two interacting systems that evolve into a state of thermal equilibrium (at constant volume).

Using \eqref{expendo} and \eqref{Psiradiuz}, the first equation of \eqref{nups} can be used to compute the viscous stress $\Pi$, 
\begin{equation}
\Pi = \dfrac{c_v^n c_v^\gamma}{c_v^n + c_v^\gamma} \bigg( \dfrac{1}{3} - \dfrac{1}{c_v^n} \dfrac{\partial P_n}{\partial \Theta_n} \bigg|_n  \bigg) (\Theta_\gamma - \Theta_n) \, .
\end{equation}
Considering that we are dealing with small deviations from equilibrium (as our goal is to match the theory of Israel and Stewart), this formula for the viscous stress can be more conveniently rewritten as 
\begin{equation}\label{secondoPiouz}
\Pi = c_v^\gamma \bigg(\dfrac{1}{3} -  \dfrac{\partial \Psi}{\partial \mathcal{U}} \bigg|_n   \bigg) (\Theta_\gamma - \Theta_n) \,  ,
\end{equation}
where we made use of the equilibrium identity \eqref{Lidentituz} proved in appendix \ref{AppendB1} (the thermodynamic derivative is performed imposing $\Psi=\Psi_{\text{eq}}$ and $\mathcal{U}=\mathcal{U}_{\text{eq}}$). This formula for $\Pi$ is equivalent to equation (33) of \citet{UdeyIsrael1982}, once one realizes that in our two-temperature model the quantity $B$ that they introduce is given by
\begin{equation}
    a_R \Theta_\gamma^4 = a_R \Theta_n^4(1+B)  \, \,  \Rightarrow  \, \,  a_R \Theta_n^4 B \approx c_v^\gamma (\Theta_\gamma - \Theta_n)
\end{equation}
As a final step, we need to impose the equivalence between the Israel-Stewart entropy production \eqref{quandoIsrproduceentrop} and our formula \eqref{quandoiorpprodent}; this leads us to the identification
\begin{equation}
\dfrac{\Pi^2}{\zeta} = \dfrac{\chi \, c_v^\gamma}{\Theta_\gamma} \,  (\Theta_n - \Theta_\gamma)^2 \, .
\end{equation}
Using \eqref{secondoPiouz} and isolating the bulk viscosity coefficient, we obtain
\begin{equation}\label{GSS}
\zeta = \dfrac{4 a_R \Theta_\gamma^4}{\chi} \bigg( \dfrac{1}{3} - \, \dfrac{\partial \Psi}{\partial \mathcal{U}} \bigg|_n   \bigg)^2,
\end{equation}
in exact agreement with the expression for the viscosity coefficient given in equation (2.42) of \citet{Weinberg1971}. 

Therefore, we have shown that Weinberg's formula for the radiation-mediated bulk viscosity coefficient $\zeta$ describes a pure two-temperature effect, where dissipation arises from the heat exchange between the matter and the radiation fluid. It is also important to note that, given the equivalence between the present UEIT model and the theory for radiation-mediated bulk viscosity of \citet{GavassinoRadiazione}, Weinberg's formula \eqref{GSS} for the bulk viscosity coefficient must also coincide with equation (142) of \citet{GavassinoRadiazione}. This is verified explicitly in appendix \ref{AppB22}.

\section{Conclusions}

In the first part of this review (Sections 1-5) we have revised the generic instability problem of relativistic Navier-Stokes \citep{Hiscock_Insatibility_first_order} from the perspective of classical field theory. The existence of unphysical gapped modes, typical of the first-order theories of dissipation, is a common feature of field theories in the slow-evolution limit. A first minimal example is the Schr\"{o}dinger dynamics (the slow limit of the Klein-Gordon model): in a boosted frame, unphysical solutions appear. 

The case of the diffusion equation is similar (as expected, since the heat equation is the Wick-rotated version of the Schr\"{o}dinger equation), with the difference that these modes have the tendency to grow with time and, therefore, they are a source of instability. 
What really drives the instability in the diffusion case, and more in general in the hydrodynamic theories of \citet{Eckart40} and \citet{landau6}, is the enforcement of the second law along these modes \citep{GavassinoLyapunov_2020}: the gapped modes give a positive contribution to the total entropy, so that they can only grow with time, leading to an instability. 
This analysis showed a difficulty in connecting a first-order hydrodynamic model for a relativistic dissipative fluid with the corresponding non-equilibrium thermodynamic description.

In the second part (starting with Section 6), we have introduced the the ideas of Unified Extended Irreversible Thermodynamics (already scattered in the literature, but presented here in a systematic way), which aims to solve the shortcomings of a first-order description, namely the connection of the phenomenological hydrodynamic model with non-equilibrium thermodynamics. 
In fact, we have discussed how every phenomenological UEIT model has, at least in principle, a natural connection with statistical mechanics and is Lyapunov-stable by construction (provided that the microscopic input makes sense). 
This connection also ensures that the gapped modes are well behaved (i.e. are expected to relax in a certain initial transient): most of UEIT models are subject to the Relaxation Effect \citep{LindblomRelaxation1996}, meaning that they exhibit a Navier-Stokes-type structure in the slow limit (without manifesting the pathologies of relativistic Navier-Stokes models).

In most non-UEIT formulations of hydrodynamics, a fluid is characterised by its fluxes ($T^{\nu \rho},s^\nu,n^\nu$) expressed in terms of some equilibrium-type thermodynamic fields ($u^\sigma,\Theta,\mu$), as well as their derivatives, via the so-called constitutive relations, and the dynamics is governed by the associated conservation laws, see e.g. \citet{kovtun_lectures_2012}. 
In these formulations, the constitutive relations are written as derivative-expansions (namely, expansions in powers of derivatives) of the fluxes. 
%
On the other hand, in UEIT, the fluid is seen as  a collection  of non-equilibrium thermodynamic systems, one for each fluid element, whose dissipative evolution is coupled to the gradients dynamically, but not directly at the level of the constitutive relations. In fact, the idea of UEIT is to enlarge the number of fields of the theory, adding to the primary equilibrium-type fields $(u^\sigma,\Theta,\mu)$ some additional non-equilibrium thermodynamic variables (e.g. non-conserved particle fractions, multiple temperatures, or even the stress tensor itself), and modelling the entropy production directly from thermodynamics. The gradients are assumed to be sufficiently small on the scale of the fluid elements not to enter into the constitutive relations (namely, individual fluid elements are approximately homogeneous), but only into the hydrodynamic equations (namely, in how different fluid elements interact with each other).  

An important formal result, discussed in Section \ref{JonasLatinista} concerns the structure of the theories of \citet{Israel_Stewart_1979} and  of \citet{carter1991}: we have shown that both are UEIT models, meaning that  they automatically inherit the desirable features of stability and well posedness that stem from the UEIT principles (when fed with a reasonable microscopic input), in agreement with~\citet{PriouCOMPAR1991}.

Finally, to show a concrete example of how the UEIT formalism works, we have re-derived the universal  model for bulk viscous fluids of \citet{BulkGavassino}. In the last section, this universal theory for bulk viscosity has been used to obtain the expression of \citet{Weinberg1971} for the  radiation-mediated bulk viscosity coefficient directly from non-equilibrium thermodynamics, without the need of solving the kinetic equation of the photon gas. This provides a thermodynamic proof for the argument of \citet{UdeyIsrael1982}, according to which radiation-mediated bulk viscosity can be seen as a pure two-temperature effect.

\section*{Acknowledgments}
Partial support comes from PHAROS, COST Action CA16214. L.G acknowledges support from Polish National Science Centre (NCN) grant OPUS 2019/33/B/ST9/00942 (B. Haskell). M.A. acknowledges support from the Polish National Science Centre (NCN) grant SONATA BIS 2015/18/E/ST9/00577 (B. Haskell). 
We thank Giovanni Camelio and Brynmor Haskell for useful discussion.

\appendix

\section{Appendix A: entropy growth along the gapped modes}\label{entropyplots}

We discuss the behaviour of the total entropy along the boost-generated gapped mode of the diffusion equation. The method is the same as the one employed in \citet{GavassinoLyapunov_2020}.
For simplicity, we work in $1+1$ dimensions and denote Alice's inertial frame by $(t,x)$ and Bob's inertial frame by $(t',x')$. The two coordinate systems are connected by the boost
\begin{equation}\label{BOOOOOOOOOOOOOst}
t' = \gamma(t-wx)   \quad \quad    x' = \gamma (x-wt) \, .
\end{equation}
We consider the first-order constitutive relations (neglecting overall additive constants)
\begin{equation}\label{constitutz}
J^\nu=
\left( \begin{matrix}
   c_v \Theta  \\
 q \\
\end{matrix} \right)
 \quad \quad  
s^\nu=
\left( \begin{matrix}
   c_v \ln \Theta  \\
 q/\Theta \\
\end{matrix} \right).
\end{equation}
Assuming the Fourier law \eqref{FFouRRieRRzxub}, we have
\begin{equation}
q=-\kappa \, \partial_x \Theta, \, 
\end{equation}
and the condition $\nabla_\nu J^\nu=0$ becomes
\begin{equation}
\partial_t \Theta - \mathcal{D} \partial_x^2 \Theta =0,
\end{equation}
which is the $1+1$ version of the diffusion equation \eqref{demifr}. 
Along the solutions of this equation the entropy production is
\begin{equation}\label{parunziov}
\nabla_\nu s^\nu = \dfrac{q^2}{\kappa \Theta^2} = \kappa \bigg( \dfrac{\partial_x \Theta}{\Theta}  \bigg)^2 ,
\end{equation}
which is strictly non-negative. 
Under the boost \eqref{BOOOOOOOOOOOOOst}, the constitutive relations \eqref{constitutz} become
\begin{equation}
J^{\nu'}= \gamma 
\left( \begin{matrix}
    c_v \Theta -w q  \\
 q - w c_v \Theta \\
\end{matrix} \right)
 \quad \quad  
s^{\nu'}= \gamma
\left( \begin{matrix}
   c_v \ln \Theta -w q/\Theta  \\
 q/\Theta - w c_v \ln \Theta  \\
\end{matrix} \right).
\end{equation}
To study the behaviour of the entropy along the spurious mode of subsection \ref{ilsuccodelproblema} we only need to impose the condition of homogeneity in Bob's reference frame. This allows us to convert the energy conservation and the second law into the ordinary differential equations
\begin{equation}
\partial_{t'} J^{t'}=0   \quad \quad    \partial_{t'} s^{t'} \geq 0 \, .
\end{equation}
The first equation implies that we can define the constant of motion
\begin{equation}
\mathcal{E}:= J^{t'} = \gamma (c_v \Theta - wq) \, ,
\end{equation}
which can be used to write $q$ as a function of $\Theta$ along the mode:
\begin{equation}
q (\Theta) = \dfrac{c_v}{w} \bigg( \Theta - \dfrac{\mathcal{E}}{ \gamma c_v}  \bigg) \, .
\end{equation}
This allows us to rewrite the entropy density as
\begin{equation}
s^{t'}(\Theta) = \gamma c_v \bigg( \ln \Theta +\dfrac{\mathcal{E}}{\gamma c_v \Theta} -1  \bigg).
\end{equation}
This function has its absolute minimum in the equilibrium state, identified by the condition
\begin{equation}
q=0   \quad    \Longleftrightarrow   \quad   \Theta = {\mathcal{E}}/({\gamma \, c_v}) \, .
\end{equation}
Therefore, the entropy density grows together with the amplitude of the mode. Hence, imposing that the second law is valid along the mode is the very origin of the growth of the mode itself with time. This originates the instability.

\section{Appendix B: Thermodynamic calculations}

In this appendix we prove some useful thermodynamic relations in the context of the model for radiation-mediated bulk viscosity presented in section \ref{TwoTemper}. 
All the calculations are performed assuming local thermodynamic equilibrium, i.e.  $\Theta_\gamma = \Theta_n = \Theta$. This condition is also used as a constraint while performing partial derivatives, reducing the number of free thermodynamic variables from 3 to 2.

\subsection{A useful identity}\label{AppendB1}

We start from the observation that (use the chain rule with $\Theta$)
\begin{equation}
\dfrac{\partial \Psi}{\partial \mathcal{U}} \bigg|_n = \dfrac{\partial \Theta}{\partial \mathcal{U}} \bigg|_n \bigg( \dfrac{\partial P_n}{\partial \Theta} \bigg|_n + \dfrac{4}{3} a_R \Theta^3    \bigg).
\end{equation}
On the other hand,
\begin{equation}
\dfrac{\partial \mathcal{U}}{\partial \Theta} \bigg|_n = c_v= c_v^n + c_v^\gamma.
\end{equation}
Hence,
\begin{equation}
\dfrac{\partial \Psi}{\partial \mathcal{U}} \bigg|_n =\dfrac{1}{c_v^n + c_v^\gamma} \bigg( \dfrac{\partial P_n}{\partial \Theta} \bigg|_n + \dfrac{c_v^\gamma}{3}   \bigg).
\end{equation}
Subtracting $1/3$ to both sides, we finally obtain
\begin{equation}\label{Lidentituz}
\dfrac{\partial \Psi}{\partial \mathcal{U}} \bigg|_n - \dfrac{1}{3} = \dfrac{c_v^n}{c_v^n + c_v^\gamma} \bigg( \dfrac{1}{c_v^n} \dfrac{\partial P_n}{\partial \Theta} \bigg|_n - \, \dfrac{1}{3} \bigg).
\end{equation}

\subsection{Chemical-like formula for Weinberg's bulk viscosity}\label{AppB22}

Equation (142) of \citet{GavassinoRadiazione} can be easily rewritten in the more convenient form
\begin{equation}\label{dalpeaperradiazio}
\zeta = \dfrac{\Theta}{\chi c_v^\gamma} \bigg( \dfrac{\partial x_{s\gamma}}{\partial v} \bigg|_{x_s} \bigg)^2.
\end{equation}  
Our task is to prove that this expression is equivalent to \eqref{GSS}.
Let us, first of all, focus on the derivative in the round brackets. Recalling the definition \eqref{xsgMama}, we immediately obtain
\begin{equation}
\dfrac{\partial x_{s\gamma}}{\partial v} \bigg|_{x_s} = c_v^\gamma \bigg( \dfrac{1}{3} + \dfrac{v}{\Theta} \dfrac{\partial \Theta}{\partial v} \bigg|_{x_s}  \bigg),
\end{equation}
which, plugged, into \eqref{dalpeaperradiazio}, gives
\begin{equation}\label{intermezzuz}
\zeta = \dfrac{c_v^\gamma \Theta}{\chi}  \bigg( \dfrac{1}{3} + \dfrac{v}{\Theta} \dfrac{\partial \Theta}{\partial v} \bigg|_{x_s}  \bigg)^2.
\end{equation}
The second step consists of deriving a useful thermodynamic identity for the second term in the round parenthesis. Imposing $\mathbb{A}^A =0$ in equations \eqref{differenz} and \eqref{LURRO}, the thermodynamic differential $d\tilde{\mathcal{U}}$ implies the usual  Maxwell relation of equilibrium thermodynamics
\begin{equation}
d \tilde{\mathcal{U}} = -\Psi dv +\Theta dx_s  
\quad   \Rightarrow   \quad
\dfrac{\partial \Theta}{\partial v} \bigg|_{x_s} \!\! =   - \dfrac{\partial \Psi}{\partial x_s} \bigg|_v 
\end{equation}
Now, applying the chain rule on the right-hand side to obtain a differentiation in $\mathcal{U}$, and recalling that $v=1/n$, we obtain the useful identity we were looking for:
\begin{equation}
\dfrac{\partial \Psi}{\partial \mathcal{U}} \bigg|_n = - \, \dfrac{v}{\Theta} \dfrac{\partial \Theta}{\partial v} \bigg|_{x_s} \, .
\end{equation}
The bulk viscosity coefficient $\zeta$ in \eqref{intermezzuz} can be finally expressed as 
\begin{equation}
\zeta = \dfrac{4 a_R \Theta^4}{\chi}  \bigg( \dfrac{1}{3} - \dfrac{\partial \Psi}{\partial \mathcal{U}} \bigg|_n \bigg)^2 \, ,
\end{equation}
which is what we wanted to prove.


\bibliography{test}

\end{document}